\documentclass[reprint,amsmath,amssymb,aps,superscriptaddress,pre]{revtex4-2}
\usepackage{graphicx}
\usepackage{color}
\usepackage{hyperref}
\usepackage{longtable}
\usepackage{booktabs}
\usepackage{siunitx}
\usepackage{bm}
\usepackage{xr}

\newcommand{\figref}[1]{Fig.~\ref{#1}}
\newcommand{\figsref}[1]{Figs.~\ref{#1}}
\renewcommand{\eqref}[1]{Eq.~(\ref{#1})}
\newcommand{\tabref}[1]{Table~\ref{#1}}
\newcommand{\secref}[1]{Section~\ref{#1}}

\newcommand{\pval}{p_{\text{val}}}

\newcommand{\joff}{j_{\text{off}}}
\newcommand{\jon}{j_{\text{on}}}
\newcommand{\tDiff}{t_{\text{Diff}}}
\newcommand{\kT}{k_{\text{B}}T}
\newcommand{\ccoex}{c_{\mathrm{coex}}}

\newcommand{\ZH}{Z_{\text{H}}}
\newcommand{\Kd}[1][]{K_{\text{d}#1}}
\newcommand{\Ld}{L_{\text{d}}}

\newcommand{\Pb}{P_{\text{b}}}
\newcommand{\nd}{n_{\text{d}}}
\newcommand{\ns}{n_{\text{s}}}

\newcommand{\dHHinteract}{d_{\text{HH,interact}}}

\newcommand{\Rgbrd}{R_{\text{g,BRD4}}}

\newcommand{\Rgbrdgas}{R_{\text{g,BRD4,gas}}}
\newcommand{\Rgtail}{R_{\text{g,tail}}}

\newcommand{\Rgi}{R_{\text{g,i}}}
\newcommand{\Rgj}{R_{\text{g,j}}}

\newcommand{\Llinker}{L_{\text{linker}}}

\newcommand{\Nnucleosomes}{N_{\text{H}}}

\newcommand{\Nac}{N_{\text{Ac}}}

\newcommand{\aHH}{a_{\text{HH}}}

\newcommand{\FTc}{F_{\text{c}}}
\newcommand{\FT}{F}

\newcommand{\FTinit}{F_{\text{init}}}
\newcommand{\FTeq}{F_{\text{eq}}}
\newcommand{\lambdamy}[1][]{\lambda_{\text{#1}}}
\newcommand{\Smy}[1][]{S_{\text{#1}}}
\newcommand{\nB}{N_{\text{B}}}
\newcommand{\Vac}{V_{\text{Ac}}}
\newcommand{\Vnonac}{V_{\text{non-Ac}}}
\newcommand{\sgmHH}{\sigma_{\text{HH}}}

\newcommand{\lambdaeq}{\lambdamy[eq]}
\newcommand{\lambdainit}{\lambdamy[init]}
\newcommand{\Ncl}{N_{\text{cl}}}
\newcommand{\DHist}{D_{\text{H}}}

\newcommand{\pAc}{p_{\text{Ac}}}
\newcommand{\pthr}{p_{\text{thr}}}

\newcommand{\tFst}{t_{\text{1st-passage}}}
\newcommand{\Vnucleus}{V_{\text{nucleus}}}

\newcommand{\Rfar}{R_{\text{far}}}
\newcommand{\Rtop}{R_{\text{top}}}
\newcommand{\rmin}{r_{\text{min}}}
\newcommand{\Rb}{R_{\text{b}}}
\newcommand{\Rg}{R_{\text{g}}}

\newcommand{\Dlangevin}{D_{\text{Langevin}}}

\newcommand{\tauLangevin}{\tau_{\text{Langevin}}}

\newcommand{\mWT}{m_{\text{BRD4,WT}}}

\newcommand{\lp}{l_{\text{p}}}

\newcommand{\lIDR}{\ell_{\text{IDR}}}
\newcommand{\nIDR}[1][]{N_{\text{IDR}#1}}
\newcommand{\lAA}{l_{\text{AA}}}

\newcommand{\Ubond}{U_{\text{bond}}}
\newcommand{\Ubind}{U_{\text{bind}}}
\newcommand{\Uwca}{U_{\text{WCA}}}
\newcommand{\Ugauss}[1][]{U_{\text{gauss}#1}}

\newcommand{\Rcut}{R_{\text{cut}}}

\newcommand{\mBD}[1]{mBD#1${}$}

\begin{document}

\title{Co-condensation and multivalency enable acetylation-sensitive, concentration-robust assembly of BRD4 condensates}
\author{Yury Polyachenko}
\affiliation{Department of Chemistry, Princeton University, Princeton, NJ 08544, USA}
\author{Hans-Frederick Watanabe}
\affiliation{Department of Chemistry, Princeton University, Princeton, NJ 08544, USA}
\author{Alexei Korolev}
\affiliation{Department of Chemistry, Princeton University, Princeton, NJ 08544, USA}
\author{William M. Jacobs}
\email{wjacobs@princeton.edu}
\affiliation{Department of Chemistry, Princeton University, Princeton, NJ 08544, USA}
\date{\today}

\begin{abstract}
Biomolecular condensates must assemble at specific locations and times inside living cells to perform their biological functions.
However, it remains unclear whether condensate formation can achieve sufficient spatiotemporal precision, responding sensitively to local chemical modifications while remaining robust to fluctuations in protein concentration.
Here we study chromatin-associated BRD4 assemblies to determine whether co-condensation with chromatin can provide the degree of discrimination required for selective targeting of highly acetylated chromatin regions.
Using an ultra-coarse-grained molecular-dynamics model, we find that heterogeneous nucleation alone does not provide robust spatial discrimination, whereas co-condensation leads to spatial selectivity and rapid formation of BRD4 assemblies across a broad range of subsaturated conditions.
Multivalent binding between BRD4 and acetylated histone tails sharpens the dependence of co-condensation on acetylation density through combinatorial effects, increasing discrimination between highly acetylated regions and weakly acetylated background chromatin.
Together, these results show how co-condensation and multivalent binding jointly enable the sensitive, concentration-robust spatiotemporal discrimination required for selective chromatin-associated assembly.
\end{abstract}

\maketitle

\section{Introduction}

Membrane-less compartments formed by biomolecular condensation contribute to spatial organization within the cell, particularly in the nucleus.
Condensates have been implicated in diverse nuclear processes, including mRNA splicing in nuclear speckles \cite{Xing1995, Xing1993, Moen2004, Brown2008}, ribosome biogenesis in nucleoli \cite{Boisvert2007, Falahati2016}, and transcriptional regulation at active genomic loci \cite{Boija2018, Sabari2018, Ma2021, Zhang2021, Lyons2023, Patil2023}.
In many of these contexts, protein-rich condensates must assemble at specific locations and times in order to perform their biological functions.
A common molecular feature of condensate-forming proteins is the combination of intrinsically disordered regions (IDRs), which promote self-attraction, with structured domains that bind specific molecular targets that can be chemically modified to impart temporal control \cite{Richter2022, Cho2018, Sabari2018}.
However, it remains unclear how these molecular features can produce condensates that localize to specific intranuclear regions with high specificity, respond sensitively to local chemical modifications, and yet remain robust to fluctuations in protein concentration.

A prominent example is the transcriptional regulator BRD4.
BRD4 contains two bromodomains (BDs) that bind acetylated histone tails \cite{Crowe2016} and a C-terminal IDR-rich region that promotes phase separation at high BRD4 concentrations \cite{Strom2024}.
Histone acetylation is enriched at transcriptionally active regions and is implicated in recruitment of RNA polymerase II \cite{Devaiah2016}.
For BRD4 condensate assembly to contribute selectively to transcriptional regulation at acetylated chromatin, BRD4 would need to assemble preferentially at highly acetylated chromatin regions while avoiding condensation elsewhere in the nucleus.
Consistent with this picture, previous work showed that acetylated chromatin can accelerate BRD4 condensate formation \cite{Strom2024}, demonstrating that BRD4--chromatin interactions can locally promote assembly.

A natural physical picture for such localized assembly is heterogeneous nucleation, in which attractive interactions with a substrate lower the nucleation barrier relative to the bulk solution \cite{Chernov1984}.
This mechanism has been invoked in prior studies of intracellular phase separation \cite{shimobayashi2021nucleation, yanagawa2024multi, erkamp2024biomolecular} and, in the context of chromatin-associated condensates, implies that chromatin acts as a substrate that locally enhances condensation \cite{Shrinivas2019, Sharma2021, Strom2024}.
However, the classical theory of heterogeneous nucleation does not readily account for selective targeting in biologically relevant scenarios where the nucleoplasmic volume exceeds that of any individual genomic locus by many orders of magnitude.
Specifically, classical nucleation theory predicts that differences between on-target (heterogeneous) nucleation and off-target (homogeneous) nucleation are greatest at small supersaturations, where the dilute BRD4 concentration is close to the saturation concentration at which the condensed and dilute phases coexist.
However, both nucleation rates vary strongly with supersaturation in this regime, rendering selective targeting extremely sensitive to BRD4 concentration \cite{porter2009phase}.
At the same time, this framework predicts that modest differences in acetylation density produce continuous changes in nucleation barriers rather than sharp thresholds, limiting discrimination between target and background chromatin.

\begin{figure*}
\centering
\includegraphics[width=0.75\textwidth]{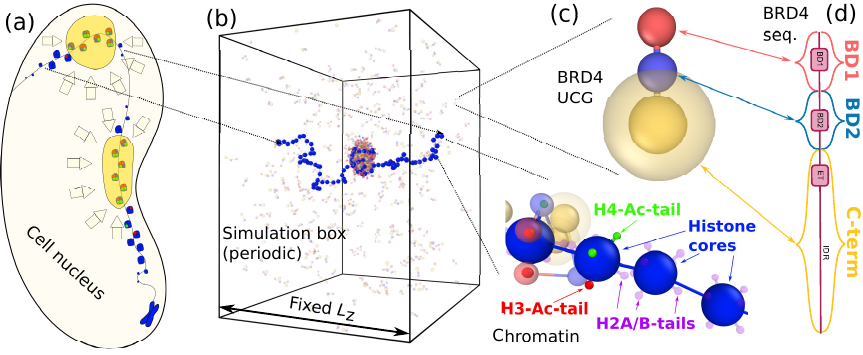}
\caption{
Overview of targeted BRD4 assembly and the ultra-coarse-grained (UCG) model.
(a)~Schematic illustration of preferential assembly of BRD4 at highly acetylated (Ac) chromatin regions. Blue circles denote nucleosomes, while red and green patches indicate acetylated H3 and H4 histone tails. Yellow shading represents condensed BRD4, whereas the light background indicates dilute protein. Arrows highlight selective localization to high-Ac regions.
(b)~Snapshot of a representative simulation configuration. In this subsaturated system, a stable, finite BRD4 assembly forms at the acetylated chromatin segment.
(c)~\textit{Top:}~UCG representation of BRD4 consisting of two bromodomain (BD) beads and one C-terminal bead representing the C-terminal IDR-rich region. Solid regions indicate steric cores, while semi-transparent regions indicate attractive interactions between disordered domains.
\textit{Bottom:}~Model representation of four nucleosomes; the left two are highly acetylated, with both H3 and H4 tails acetylated, whereas non-acetylated tails (purple) interact only sterically. One BRD4 molecule is shown bound monovalently to an H3 tail through BD1, whereas the second is bound bivalently to an H3 tail through BD1 and an H4 tail through BD2. See \figsref{fig:1S5} and \ref{fig:1S2} for model parametrization.
(d)~Schematic correspondence between the BRD4 sequence and its representation in the UCG model.
}
\label{fig:1}
\end{figure*}

Here we ask whether these requirements---selective assembly that is sensitive to acetylation yet robust to BRD4 concentration fluctuations---can instead be satisfied by co-condensation below the bulk coexistence concentration, building on prior studies of polymer-assisted condensation and related co-condensation phenomena \cite{Sommer2022,Quail2021,Renger2022,Rouches2025}.
In this regime, BRD4 binding and self-attraction cooperatively drive local chromatin compaction while stabilizing finite BRD4 assemblies, so that chromatin reorganization and BRD4 enrichment occur in a coupled manner.
Because bulk phase separation is thermodynamically unfavorable below the BRD4 coexistence concentration, off-chromatin condensation is intrinsically suppressed in this regime.
Crucially, we ask whether this mechanism can discriminate highly acetylated target regions from background chromatin with lower but nonzero acetylation, and whether this discrimination remains robust to variations in BRD4 concentration.

In this Article, we use an ultra-coarse-grained (UCG) molecular dynamics (MD) model of BRD4 and chromatin to demonstrate that co-condensation can indeed satisfy these quantitative requirements for selective chromatin targeting.
We (i) characterize the concentration and acetylation-density dependence of finite BRD4 assemblies below bulk phase coexistence, (ii) quantify how BRD4 multivalency sharpens discrimination between different acetylation densities, (iii) characterize BRD4 assembly-formation kinetics to assess their sensitivity to BRD4 concentration and acetylation density, and (iv) determine the parameter regimes in which discrimination between highly acetylated target regions and background chromatin remains robust to BRD4 concentration fluctuations.
Together, these analyses show how the physical features of BRD4 and chromatin can produce selective, rapid, and concentration-robust targeting under biologically relevant conditions.

\section{Results}
\label{sec:Results}

\subsection{Coarse-grained model of BRD4 and acetylated chromatin}
\label{subsec:CGmodel}

\begin{figure*}
\centering
\includegraphics[width=0.75\textwidth]{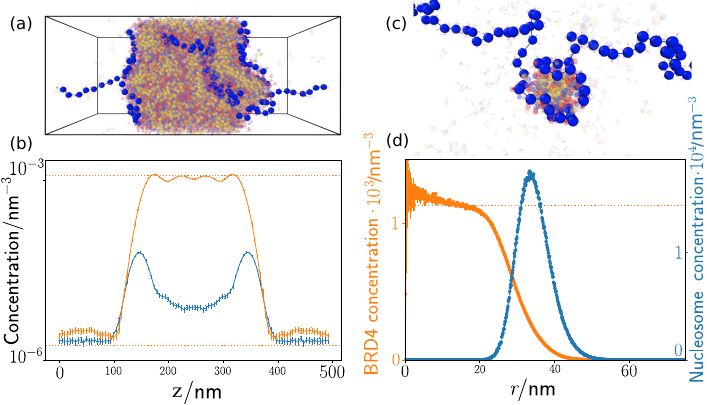}
\caption{Chromatin localizes to the interface of the BRD4-rich phase for both intermediate acetylation and finite assembly size.
(a)~Coexistence between bulk dilute and condensed BRD4 phases in the presence of chromatin at 50\% acetylation probability, $\pAc = 0.5$, and (b)~corresponding BRD4 and nucleosome concentration profiles along the $z$ direction. Chromatin is excluded from the interior of the condensed phase and instead accumulates at the interface. Dotted lines indicate dilute- and condensed-phase BRD4 concentrations measured in the absence of chromatin (Fig.~\ref{fig:1S3}); the similarity between dilute-phase BRD4 and nucleosome concentrations is coincidental. Oscillations in the condensed-phase BRD4 concentration reflect layering arising from favorable interactions between C-terminal beads.
(c)~A finite BRD4 assembly formed on an acetylated chromatin segment under subsaturated conditions with 50\% acetylation probability, and (d)~corresponding radial concentration profiles, showing that chromatin is localized near the assembly interface rather than in the interior at equilibrium. The dotted line indicates the condensed-phase BRD4 concentration in the absence of chromatin, as in (b).
See Fig.~\ref{fig:2S0} for results at other acetylation levels.}
\label{fig:2}
\end{figure*}

To test this proposed mechanism, we develop a UCG MD model of BRD4 and chromatin that reproduces the minimal ingredients required for assembly: excluded-volume interactions, specific 1:1 binding between bromodomains and acetylated histone tails, attraction between disordered regions, and chromatin deformability (\figref{fig:1}; implementation details are provided in SI~\secref{sec:model} and \figsref{fig:1S5} and \ref{fig:1S2}).
Each BRD4 molecule is represented by three beads, with two beads corresponding to the BDs and one C-terminal bead representing the C-terminal IDR-rich region.
IDR-mediated attractions, which are strongest for the C-terminal beads, primarily drive phase separation, whereas BD beads bind acetylated histone tails with 1:1 specificity.
Excluded-volume interactions set the condensed-phase concentration and prevent unphysical overlap.
This representation preserves BRD4’s anisotropic geometry and the 1:1 binding constraint required to capture multivalent interactions, while enabling simulations over system sizes and timescales relevant for assembly.

Chromatin is modeled as a semiflexible polymer of nucleosomes connected by linker DNA.
Each nucleosome contains histone tails that can be either acetylated or non-acetylated.
Acetylated tails bind BDs, whereas non-acetylated tails interact only sterically.
Histone-tail attachment points are arranged according to the approximate nucleosome geometry~\cite{Richmond1984}, with weak angular constraints used to preserve their spatial organization while retaining tail flexibility.
In simulations with chromatin, a single chromatin fiber spans the simulation box under periodic boundary conditions (\figref{fig:1}b); the box length in this direction ($z$) is held fixed so that chromatin tension can be controlled systematically (see SI~\secref{sec:simulations}).

We select acetylation patterning regimes in simulations to probe biologically relevant conditions for chromatin-associated condensate formation.
Acetylated chromatin regions in human cells span a wide range of genomic lengths under physiological conditions \cite{Panigrahi2021, Wu2023}.
Enhancer-scale regions of $\sim 1$--\SI{6}{kbp} correspond to approximately $5$--$30$ consecutive nucleosomes, depending on linker length.
Background acetylation levels near active loci are on the order of $10$--$15\%$, whereas highly acetylated regions can reach $70$--$80\%$ \cite{Litt2001}.
We therefore choose $\Nac = 20$ consecutive nucleosomes as a representative acetylated segment length in the main text and compare to $\Nac = 30$ in the Supplementary Information.
Within this segment, a fixed fraction $\pAc$ of the H3 and H4 histone tails is acetylated, with the acetylated sites randomly distributed along the segment.
Thus, varying $\pAc$ changes the density of available acetylated binding sites while the BD--Ac interaction strengths are held fixed.
Exploring a broad range of $\pAc$ allows us to assess whether assembly discriminates between modest background acetylation and strongly acetylated domains.

\subsection{Finite BRD4 assemblies are distinct from bulk condensates}
\label{subsec:finiteSize}

We first examine how chromatin partitioning differs between two contrasting regimes: bulk phase coexistence and finite assemblies formed under subsaturated conditions.
We define the saturation as
\begin{equation*}
  S \equiv \frac{c_{\rm far}}{\ccoex},
\end{equation*}
where $c_{\rm far}$ is the far-field BRD4 concentration and $\ccoex$ is the dilute-phase concentration at bulk coexistence in the absence of chromatin.
Thus, finite assemblies formed below bulk coexistence are characterized by $S<1$.
By contrast, in an equilibrated two-phase system the far-field dilute-phase concentration is $\ccoex$, so $S=1$; we therefore refer to this regime simply as bulk phase coexistence.
Baseline BRD4 phase behavior without chromatin, from which $\ccoex$ is determined, is shown in Fig.~\ref{fig:1S3}.

Chromatin perturbs phase equilibrium only weakly in the case of bulk phase separation.
In this regime, we characterize the equilibrium behavior using constant-volume ($NVT$) direct-coexistence simulations, in which the dilute and condensed phases are separated by an approximately planar interface (\figref{fig:2}a).
The dilute- and condensed-phase BRD4 concentrations remain close to the coexistence concentrations in the absence of chromatin, indicating that chromatin does not significantly alter the bulk phase behavior (\figsref{fig:2}a,b and \ref{fig:2S0}).
Instead, chromatin redistributes within the coexisting phases depending on acetylation.
At low acetylation, chromatin is preferentially excluded from the condensed phase due to dominant excluded-volume interactions, whereas at high acetylation it is preferentially incorporated into the condensed phase through BD binding.
At intermediate acetylation, chromatin tends to localize near the interface between the dilute and condensed phases (\figref{fig:2}a,b).
Partitioning to the interface occurs because 1:1 binding interactions between acetylated chromatin and BDs attract chromatin to the condensate but are insufficiently numerous to compensate for the disruption of cohesive BRD4 interactions within the condensed phase.
Thus, in the bulk coexistence regime, chromatin partitioning is strongly dependent on acetylation density.

\begin{figure*}
    \centering
    \includegraphics[width=\textwidth]{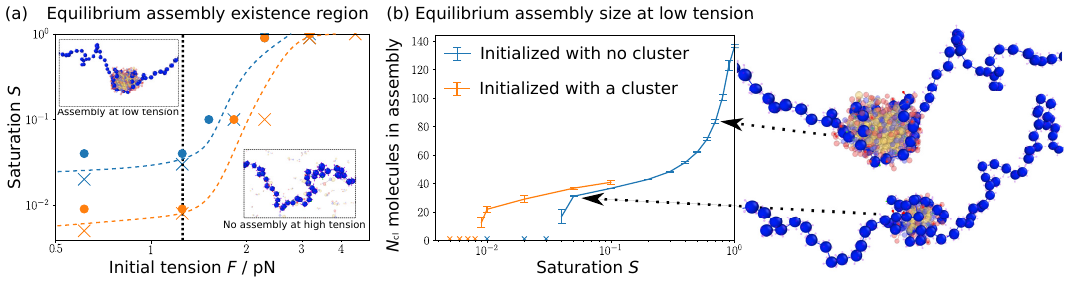}
    \caption{Chromatin tension and BRD4 saturation regulate BRD4 assembly formation and stability. Behavior is shown for a fully acetylated region ($\pAc = 1$) with $\Nac = 20$.
      (a)~Assembly stability diagram showing the presence or absence of stable BRD4 assemblies as a function of chromatin tension $\FT$ and BRD4 saturation $S$. Blue filled circles indicate formation of a stable assembly from a no-cluster initial condition, while blue crosses indicate conditions where no assembly is observed. Orange filled circles indicate that a preformed BRD4 cluster leads to a stable assembly at equilibrium, while orange crosses indicate conditions where preformed clusters dissolve. The separation between assembly formation (blue) and dissolution (orange) boundaries indicates hysteresis; dashed lines are guides to the eye.
      (b)~Equilibrium cluster size $\Ncl$ of the BRD4 assembly as a function of $S$ at a sufficiently low tension $F \approx \SI{1.26}{pN}$, corresponding to the vertical dotted line in (a). Hysteresis between formation and dissolution protocols is observed at low $S$. Snapshots illustrate representative equilibrium configurations at two values of $S$.
    }
    \label{fig:3}
\end{figure*} 

Qualitatively different behavior emerges under subsaturated conditions ($S<1$).
In this regime, we maintain a constant far-field BRD4 concentration by performing simulations in an $NP_xP_yL_zT$ ensemble.
The pressures in the $x$ and $y$ directions are controlled while the box length $L_z$ is fixed, allowing the far-field BRD4 concentration and chromatin extension to be varied independently.
For the conditions considered here, the dilute BRD4 phase is approximately ideal, so we impose $P_x=P_y=S\ccoex k_{\mathrm{B}}T$, resulting in a far-field concentration $c_{\rm far}\approx S\ccoex$ at equilibrium (\figref{fig:npxpy_validation}).
Chromatin extension is controlled by the fixed box length $L_z$.
Although $P_z$ is not externally controlled, the far-field BRD4 gas equilibrates to an isotropic pressure tensor.
Further details and direct validation of the far-field concentration control are provided in SI~\secref{sec:npxpylzt}.

Under these conditions, acetylated chromatin stabilizes finite BRD4 assemblies, which form only on acetylated chromatin segments (\figref{fig:2}c,d).
These assemblies exhibit interior BRD4 concentrations comparable to the bulk condensed phase, despite forming below coexistence.
The acetylated chromatin segments collapse together with the assembling BRD4 molecules to maximize 1:1 binding with BDs.
This mechanism, which we refer to as co-condensation~\cite{Quail2021, Renger2022}, corresponds to the coupled reorganization of chromatin and accumulation of BRD4 into a finite assembly stabilized by both binding and self-attraction.
Unlike in the bulk coexistence regime, this interfacial localization is robust with respect to acetylation density when the acetylated region is limited to a biologically relevant stretch of $\Nac \approx 20$ nucleosomes (\figref{fig:2S0}).

The partitioning behavior of chromatin in both regimes is strongly influenced by the organization of BRD4 molecules in the condensed phase.
In bulk condensates, segregation between BD-rich and C-terminal-rich regions due to the anisotropic geometry and different interaction strengths results in local concentration variations (\figref{fig:2}b).
BD-rich regions are found throughout the interior of bulk condensates, allowing chromatin to be incorporated at sufficiently high acetylation.
By contrast, in finite assemblies, C-terminal beads dominate the interior due to their favorable self-interactions, while BD beads are enriched near the interface.
Because acetylated chromatin binds specifically to BD beads, it preferentially associates with these interfacial regions at equilibrium.
Crossover between these two regimes can be observed by increasing $\Nac$, which causes the equilibrium BRD4 assembly size to grow and internal BD-rich regions to emerge, leading to increased chromatin incorporation.
Taken together, these results show that chromatin partitioning in both regimes is governed by the common principle of maximizing BD binding to acetylated chromatin while minimizing disruption of the preferred condensed-phase organization of BRD4.

\subsection{Conditions for BRD4--chromatin co-condensation and finite assembly stability}
\label{subsec:stability}

We next characterize the conditions for equilibrium co-condensation of $\Nac = 20$ acetylated regions in this model.
Using the $NP_xP_yL_zT$ ensemble, we systematically vary both the saturation $S$ and the tension on the chromatin fiber.
The initial chromatin tension in the simulation is related to the chromatin extension parameter $\lambda \equiv \Nnucleosomes \aHH / L_z$, where $\Nnucleosomes$ is the number of nucleosomes and $\aHH$ is the equilibrium spacing between neighboring nucleosomes.
Large $\lambda$ corresponds to low initial tension and permits substantial local chromatin compaction, whereas $\lambda \approx 1$ corresponds to a nearly straight, high-tension fiber.
Independent simulations without BRD4 establish the mapping between $\lambda$ and the resulting mechanical tension $\FT$ (\figref{fig:3S1}a).

Co-condensation occurs when the free-energy change due to BRD4 adsorption to chromatin, including both BD binding and IDR-mediated self-attraction, exceeds the mechanical and entropic cost of bending and compacting the chromatin fiber.
The balance of these factors depends on both BRD4 saturation and chromatin tension, and can be understood by examining the equilibrium behavior of a fully acetylated ($\pAc = 1$) $\Nac = 20$ segment (\figsref{fig:3}a and \ref{fig:3S1}b,c; see also \figref{fig:3S2} for $\Nac = 30$).
Stable BRD4 assemblies form only when chromatin tension lies below a critical threshold $\FTc \approx 1.5$--$\SI{2}{pN}$, which is comparable to tensions reported in single-molecule experiments probing condensate interactions with DNA \cite{Renger2022, Quail2021}.
Below this threshold, the minimum far-field BRD4 concentration required for stable assembly is approximately two orders of magnitude below the bulk coexistence concentration.
However, results obtained from simulations initialized with and without a pre-formed assembly define distinct formation and dissolution boundaries, reflecting hysteresis associated with a finite free-energy barrier (\figref{fig:3}a).
Above the critical tension threshold, chromatin remains extended and assembly is observed only very near coexistence or under supersaturated conditions, where bulk phase separation becomes thermodynamically favorable.

BRD4 saturation also controls the equilibrium size of finite assemblies below the critical tension threshold (\figref{fig:3}b).
The equilibrium size is strongly influenced by BD binding, which is limited by the total number of acetylated histone tails ($4 \pAc \Nac = 80$ for a fully acetylated $\Nac = 20$ segment).
Just above the minimum saturation for co-condensation, assemblies contain $\Ncl \approx 30$--$40$ BRD4 molecules, consistent with most molecules engaging both BDs in binding to distinct acetylated histone tails.
As saturation increases, the entropic penalty for BRD4 adsorption decreases and assembly sizes approach the single-binding limit, $\Ncl \approx 4 \pAc \Nac$.
However, near coexistence ($S \gtrsim 0.8$), the equilibrium assembly size increases sharply.
This behavior reflects the metastability of micelle-like BRD4 clusters in the absence of chromatin within this saturation range (\figref{fig:2S2}); in this relatively narrow regime, acetylated chromatin stabilizes clusters that would otherwise exist only transiently.

These results show that finite BRD4 assemblies are governed by the interplay among BRD4 self-attraction, multivalent binding to acetylated chromatin, entropic forces, and chromatin mechanics.
Co-condensation requires both sufficiently low chromatin tension and sufficiently high BRD4 concentration to overcome mechanical and entropic costs, defining a regime in which stable, finite assemblies are supported by BD--chromatin interactions without undergoing bulk phase separation.

\begin{figure*}
    \centering
    \includegraphics[width=\textwidth]{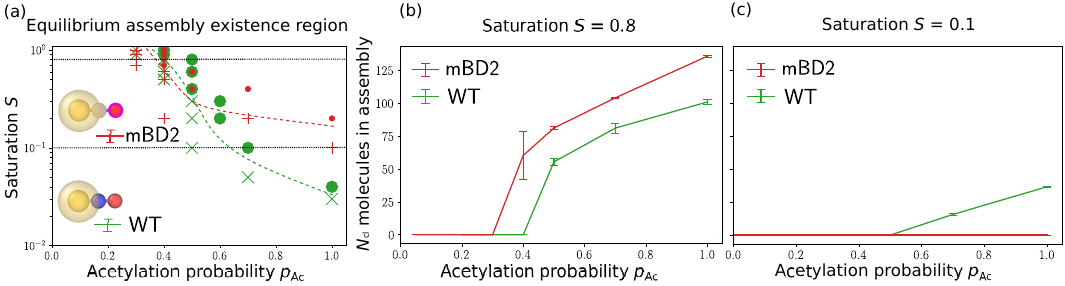}
    \caption{BRD4 multivalency sharpens the dependence of co-condensation on acetylation density.
      (a)~Co-condensation as a function of saturation $S$ and the acetylation probability $\pAc$ of H3 and H4 histone tails at low chromatin tension ($F \approx \SI{1.26}{pN}$) with $\Nac = 20$. Filled circles indicate that a stable BRD4 assembly forms from a no-cluster initial condition, while crosses indicate that no assembly forms within the simulation time. Green symbols correspond to the wild-type (WT) BRD4 model, whereas red symbols correspond to the single-active-domain \mBD2 model, in which BD2--Ac binding is disabled while the effective BRD4--histone binding affinity is approximately matched at intermediate $\pAc$ (see \figref{fig:4S3}). WT BRD4 and \mBD2 both co-condense with acetylated chromatin at sufficiently high $\pAc$, but WT BRD4 exhibits a sharper dependence on acetylation density. Dashed lines are guides to the eye. Black dotted lines indicate the two values of $S$ used in (b) and (c). Insets illustrate UCG representations of WT BRD4 and the \mBD2 variant.
      (b)~Equilibrium BRD4 assembly size $\Ncl$ as a function of acetylation probability $\pAc$ at constant saturation $S = 0.8$.
      (c)~$\Ncl$ as a function of acetylation probability $\pAc$ at constant saturation $S = 0.1$. WT BRD4 co-condenses with acetylated chromatin above a threshold $\pAc$, whereas \mBD2 remains dispersed across the full range of $\pAc$. The vertical axis is the same as in (b).
    }
    \label{fig:4}
\end{figure*}

\subsection{BRD4 multivalency sharpens acetylation dependence of co-condensation}
\label{subsec:AcAndMultivalent}

We now examine how BRD4 multivalency influences the sensitivity of co-condensation to acetylation density.
Throughout this analysis, chromatin tension is held fixed below the critical value at $\FT \approx\SI{1.26}{pN}$ so that assembly can proceed via co-condensation.
We then vary the acetylation probability $\pAc$ along a segment of $\Nac = 20$ nucleosomes.

At this low tension, stable assemblies form only above a threshold acetylation probability $\pthr$, with co-condensation occurring well below bulk coexistence when $\pAc > \pthr$.
This sharp dependence of co-condensation on $\pAc$ enables strong discrimination between highly acetylated regions ($\pAc \sim 0.7$--$0.8$) and chromatin with modest background acetylation levels ($\pAc \sim 0.1$--$0.15$) \cite{Litt2001}.
Importantly, for wild-type (WT) BRD4, $\pthr$ lies between $\pAc \approx 0.4$ and $0.6$ over approximately an order of magnitude variation in the far-field BRD4 concentration below coexistence (\figref{fig:4}a).
The weak dependence of $\pthr$ on $S$ indicates that this discrimination remains robust to variations in BRD4 concentration when $S \gtrsim 0.1$.

To isolate the role of multivalency in co-condensation, we compare WT BRD4, which has two active Ac-binding BDs, with a single-active-domain variant (\mBD2). In the \mBD2 model, both BD beads are retained sterically, but BD2--Ac binding is disabled and the BD1--Ac interaction strength is modestly adjusted to approximately match the effective BRD4--histone binding affinity of WT at intermediate acetylation ($\pAc \approx 0.5$; see \figref{fig:4S3}).
Although both models can co-condense with acetylated chromatin at sufficiently high $\pAc$ and $S$, their acetylation dependence differs in how the threshold behavior varies with saturation (\figref{fig:4}a).
In particular, \mBD2 fails to form stable assemblies at any acetylation level when $S \lesssim 0.1$, and its threshold acetylation probability $\pthr$ varies more strongly with $S$ at higher saturations where co-condensation can occur.
Thus, multivalency does not simply shift the threshold for co-condensation but instead sharpens its dependence on acetylation density.

These differences in thresholding behavior arise from combinatorial binding effects \cite{Curk2017,Xie2025,Xia2026}.
Because each nucleosome contains multiple acetylatable histone tails, the typical number of distinct configurations in which a BRD4 molecule can bind a nucleosome increases with acetylation density.
The number of binding configurations grows more rapidly for multivalent BRD4 than for the single-active-domain \mBD2 variant, so that the favorable free-energy change due to binding varies more steeply with $\pAc$ for WT BRD4 and reduces the sensitivity of the onset of co-condensation to BRD4 saturation.
This mechanism is supported by a combinatorial binding model and by simulations at high chromatin tension that isolate BRD4--histone interactions from co-condensation (see SI~\secref{sec:multivalency} and \figref{fig:4S1}).

Consistent with this mechanism, increasing the number of active BDs from one to two roughly doubles the slope of the co-condensation boundary when plotted as $\ln S$ versus acetylation probability for $S \gtrsim 0.1$ (\figref{fig:4}a and \figref{fig:4S2}).
Multivalency therefore sharpens the boundary between co-condensed and non-assembling regimes and expands the range of subsaturated BRD4 concentrations over which co-condensation occurs.
These effects enhance contrast between highly acetylated and weakly acetylated regions while increasing robustness to variations in BRD4 concentration.

\begin{figure*}
    \centering
    \includegraphics[width=\textwidth]{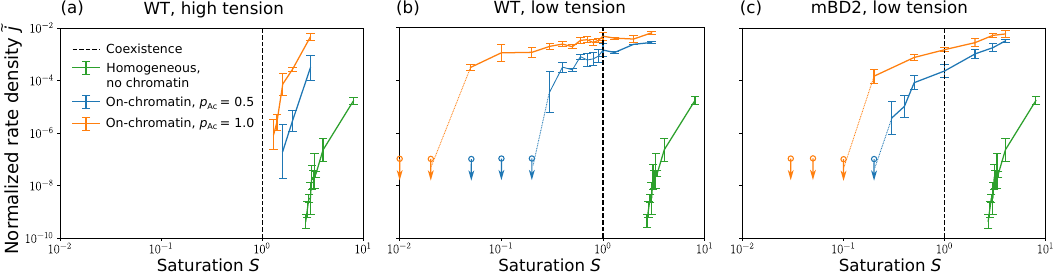}
    \caption{Co-condensation alters the kinetics of BRD4 assembly formation and reduces sensitivity to saturation.
      (a)~Nucleation rate density for WT BRD4 at high chromatin tension ($F \approx \SI{4.5}{pN}$), where co-condensation is suppressed. The homogeneous off-chromatin rate is compared with excess chromatin-associated rate densities for $\pAc=0.5$ and $\pAc=1$. Assembly occurs only for $S > 1$, and the rate increases with both saturation $S$ and acetylation probability $\pAc$.
      (b)~Assembly formation rate density for WT BRD4 at low chromatin tension ($F \approx \SI{0.6}{pN}$), where co-condensation is permitted. Stable assemblies form over a wide range of subsaturated conditions ($S < 1$), and formation rates vary weakly with $S$ at sufficiently high acetylation.
      (c)~Same as (b), but for the single-active-domain \mBD2 model. Assembly is more strongly suppressed at low saturation compared to WT BRD4.
      Open circles and arrows indicate an estimated upper bound on the formation rate under conditions where co-condensation is stable when initialized from a BRD4 cluster but assembly formation was not observed within the simulation duration.
      All calculations assume $\Nac = 20$. See \figref{fig:5S2} for $\Nac = 30$.
    }
\label{fig:5}
\end{figure*} 

\subsection{Co-condensation enables rapid assembly over a wide range of BRD4 concentrations}
\label{subsec:rates}

To understand how co-condensation modifies the kinetics of BRD4 assembly, we compare assembly formation rates above and below coexistence to the predictions of classical nucleation theory (CNT).
We measure the rate density $j$ for finite assembly formation ($S<1$) or condensed-phase nucleation ($S>1$) as a function of $S$ and $\pAc$.
We then define the dimensionless diffusion-normalized rate density $\tilde{j}=j\,t_{\rm Diff}(S)/(\ccoex S^{2/3})$ to compare rates across conditions, where $t_{\rm Diff}(S)=(2R_{\rm g})^2/[6D(S)]$ is the characteristic time for a BRD4 molecule to traverse its gyration diameter (\figref{fig:5S1}).
This normalization accounts for both the saturation dependence of molecular diffusion and the expected concentration dependence associated with diffusion over the mean intermolecular spacing.
With this normalization, CNT predicts a strongly saturation-dependent nucleation barrier near coexistence, such that the diffusion-normalized rate obeys $\ln \tilde{j}=A-B/(\ln S)^2$, with $S>1$ and $B>0$ \cite{porter2009phase}.

We perform all kinetic simulations in the $NP_xP_yL_zT$ ensemble so that depletion of the far-field BRD4 reservoir does not appreciably influence the measured formation rates.
For nucleation calculations at $S>1$, $S$ denotes the supersaturation of the initially homogeneous BRD4 phase in which $c_{\rm far}=c=S\ccoex$ prior to nucleation; following nucleation, growth of a post-critical cluster would ultimately lead to macroscopic phase separation.
By contrast, under subsaturated conditions, $S$ describes both the initial and equilibrium far-field concentration since co-condensation leads to a finite BRD4 assembly.

Classical nucleation behavior is observed for homogeneous BRD4 phase separation in the absence of chromatin and for heterogeneous nucleation in the presence of high-tension chromatin, where co-condensation is suppressed (\figref{fig:5}a).
With $S>1$, we extract the rate from the mean first-passage time to exceed a cluster size corresponding to post-critical growth.
We then compute the rate density $j$ using the appropriate normalization volume for the process being measured; for chromatin-containing simulations, the reported excess chromatin-associated rate density is obtained after accounting for the expected homogeneous contribution in the accessible volume (see SI~\secref{sec:kinetics}).
In this high-tension regime, chromatin acts as a rigid substrate that lowers the heterogeneous nucleation barrier but does not alter the underlying CNT mechanism.
Accordingly, the heterogeneous nucleation rate increases strongly with saturation and acetylation density, and its dependence on $S$ follows the CNT scaling expected for barrier-limited nucleation.
Thus, high-tension chromatin can enhance nucleation locally, but the assembly kinetics exhibit extreme sensitivity to the bulk BRD4 concentration.

At low chromatin tension, where co-condensation is permitted, assembly proceeds via a qualitatively different mechanism (\figref{fig:5}b).
We define the total assembly formation rate as $J=1/\langle\tFst\rangle$, where $\tFst$ is the earliest time at which the assembly size exceeds the threshold used to identify the formation of a stable assembly; $J$ is then converted to a rate density $j$ for a target region containing 20 acetylated nucleosomes (see SI~\secref{sec:kinetics}).
Stable assemblies form over a broad range of subsaturations, and formation rates vary only weakly with $S$ once acetylation exceeds the threshold required for stability.
In contrast to nucleation-limited dynamics, assembly does not require the formation of a large critical assembly whose size depends strongly on concentration.
Instead, BRD4 molecules adsorb onto chromatin and reorganize together with the polymer into a stable assembly, with the rate primarily limited by the diffusion of dilute BRD4 to the acetylated region over a broad range of subsaturations.
Consistent with this picture, the diffusion-normalized rate densities $\tilde{j}$ are nearly constant across saturation conditions in this regime, indicating that assembly kinetics are set primarily by transport rather than by a large nucleation barrier.
Co-condensation therefore tends to decouple rapid assembly formation rates from BRD4 saturation.

The influence of multivalency on kinetics parallels its effect on the equilibrium stability of BRD4 assemblies.
Comparison with the single-active-domain \mBD2 variant (\figref{fig:5}c) shows that multivalency extends the regime of rapid assembly to lower saturation and reduces the residual dependence of assembly formation rates on saturation above $\pthr$.
While both models exhibit accelerated assembly at high $\pAc$, \mBD2 assembles over a narrower subsaturated regime and exhibits slower assembly at lower $S$, consistent with the reduced stability of co-condensation at low $S$ in the equilibrium analysis.
Thus, multivalent binding enhances not only the thermodynamic robustness of co-condensation but also its kinetic accessibility.

\begin{figure*}
    \centering
    \includegraphics[width=\textwidth]{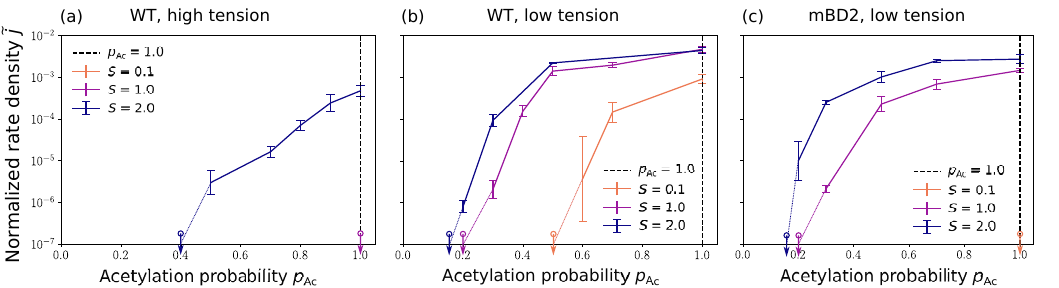}
    \caption{Assembly formation rates depend on acetylation density and chromatin tension.
      (a)~Nucleation rate density for WT BRD4 at high chromatin tension ($F \approx \SI{4.5}{pN}$), where chromatin compaction is suppressed. Rates increase steadily with acetylation probability $\pAc$.
      (b)~Assembly formation rate density for WT BRD4 at low chromatin tension ($F \approx \SI{0.6}{pN}$), where co-condensation is permitted. At sufficiently high acetylation, rates exhibit a plateau as a function of $\pAc$ at both subsaturated and supersaturated conditions.
      (c)~Same as (b), but for the single-active-domain \mBD2 model. Compared to WT BRD4, assembly formation is strongly suppressed at low saturation, while the rate plateau is less pronounced at high $\pAc$.
      As in \figref{fig:5}, open circles and arrows indicate an estimated upper bound on the formation rate.
      All calculations assume $\Nac = 20$. See \figref{fig:6S1} for simulation results with $\Nac = 30$.}
    \label{fig:6}
\end{figure*} 

The dependence of assembly rates on acetylation density further reinforces this connection between the kinetics and stability of co-condensation (\figref{fig:6}).
At high chromatin tension (\figref{fig:6}a), rates increase steadily with $\pAc$ when $S>1$, reflecting progressive lowering of the heterogeneous nucleation barrier.
By contrast, at low tension (\figref{fig:6}b), rates increase sharply once $\pAc$ exceeds the threshold for stable co-condensation and then plateau across a wide range of saturations, including deeply subsaturated conditions.
This plateau reflects approximately diffusion-limited assembly and mirrors the thresholding behavior observed in equilibrium.
For the single-active-domain \mBD2 model (\figref{fig:6}c), the plateau is weaker and assembly rates are more strongly suppressed at low $S$, again consistent with \mBD2's co-condensation behavior in equilibrium.

To estimate absolute assembly times, we calibrate the simulation dynamics using the experimentally measured intranuclear diffusion coefficient of freely diffusing BRD4 (see SI~\secref{sec:time_calibration} and \secref{sec:absolute_timescales}).
This mapping gives a characteristic BRD4 diffusion time of approximately \SI{4.4}{\mu s} over one molecular diameter.
The plateau co-condensation rates in \figsref{fig:5} and \ref{fig:6} then correspond to a mean assembly-formation time of approximately $\SI{20}{ms} \times S^{-2/3}$ for a target region containing 20 acetylated nucleosomes.
At $S=0.1$, this yields a mean assembly-formation time of approximately \SI{90}{ms}.
Thus, stable BRD4 assemblies are predicted to form on subsecond timescales even under strongly subsaturated conditions.
Although a direct quantitative comparison is complicated by differences in the number of available target sites and experimental detection thresholds, this predicted timescale is compatible with the rapid, seconds-scale formation of BRD4 condensates observed in U2OS cells \cite{Strom2024}.

Taken together, these results connect the kinetics of co-condensation to the same physical ingredients that determine its equilibrium behavior.
Multivalency sharpens acetylation-dependent thresholds and expands the range of conditions under which co-condensation can occur, resulting in approximately diffusion-limited assembly that is not controlled by a large nucleation barrier.
Consequently, co-condensation supports rapid assembly in the same parameter regime that enables strong discrimination between target and background chromatin while remaining robust to variations in BRD4 concentration.

\subsection{Conditions for selective and concentration-robust discrimination of acetylated chromatin}
\label{subsec:targetingMap}

\begin{figure*}
    \centering
    \includegraphics[width=\textwidth]{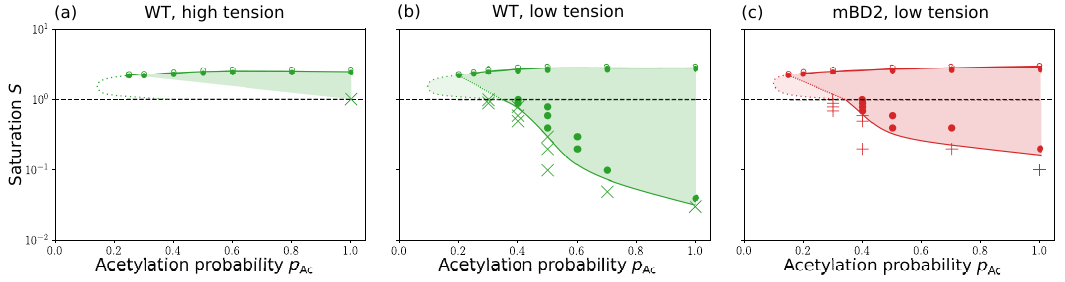}
    \caption{Parameter regimes that enable selective and concentration-robust targeting of BRD4 assemblies, shown as functions of acetylation density and saturation.
      (a)~Formation of BRD4 assemblies in the $(\pAc,S)$ plane at high chromatin tension ($F \approx \SI{4.5}{pN}$), where co-condensation is suppressed. For $S>1$, open symbols indicate $\jon v / \joff \Vnucleus = 1$, and filled symbols indicate $\jon v / \joff \Vnucleus = 10$, assuming $v / \Vnucleus = 10^{-6}$. The shaded region is a guide to the eye delineating conditions under which selective targeting is achieved ($\jon v / \joff \Vnucleus \ge 10$); the dashed line indicates that selective targeting must vanish at sufficiently low $\pAc$, although direct simulation is not feasible in this regime.
      (b)~Same analysis as in (a), but at low chromatin tension ($F \approx \SI{0.6}{pN}$), where co-condensation is permitted. For $S<1$, filled symbols indicate conditions under which BRD4 and acetylated chromatin rapidly co-condense, while crosses indicate conditions where assembly is not observed. In the subsaturated shaded region, $\jon v / \joff \Vnucleus \rightarrow \infty$ because off-target nucleation is thermodynamically suppressed.
      (c)~Same as (b), but for the single-active-domain \mBD2 model. The subsaturated region of stable assembly formation is reduced compared to WT BRD4.
      All calculations assume $\Nac = 20$ acetylated nucleosomes.}
    \label{fig:7}
\end{figure*}

We finally combine these equilibrium and kinetic analyses to identify the parameter regimes in which BRD4 can strongly discriminate between highly acetylated target regions and background chromatin while remaining robust to variations in BRD4 concentration and assembling on biologically relevant timescales.
For a single acetylated chromatin region embedded in a nucleus of accessible volume $\Vnucleus$, preferential targeting requires that the total target-associated formation rate exceed the total off-target rate.
Using $\jon$ to denote the excess chromatin-associated rate density above the homogeneous background rate density $\joff$, the target-associated rate is $(\jon+\joff)v$, where $v$ is the effective target volume.
Thus, preferential targeting requires
\begin{equation*}
  (\jon+\joff)v \gg \joff(\Vnucleus-v).
\end{equation*}
For $\Vnucleus / v \gg 1$, this condition reduces to
\begin{equation*}
  \jon v \gg \joff\Vnucleus,
\end{equation*}
which is the targeting criterion used below.
For a nucleus of radius $\sim \SI{6}{\mu m}$ and a single submicron-scale acetylated region, the ratio $\Vnucleus/v$ can approach $10^6$.
Thus, in a supersaturated system, heterogeneous nucleation would require orders-of-magnitude enhancement of the local rate to overcome the larger off-target volume.
At the same time, $\jon v$ must be sufficiently large in absolute terms to ensure rapid assembly following changes in acetylation.

These requirements are difficult to satisfy under high chromatin tension, where co-condensation is suppressed and assembly proceeds via classical heterogeneous nucleation (\figref{fig:7}a).
Because nucleation rates depend steeply on supersaturation, selective targeting occurs only in a small interval slightly above $S=1$.
Near-coexistence nucleation is slow everywhere, whereas at larger $S$ off-target nucleation becomes competitive.
Moreover, the dependence of assembly on acetylation density in this regime is continuous rather than threshold-like, so modest background acetylation can promote nucleation at sufficiently high saturation.
This narrow regime of sensitivity implies that targeting is not robust to fluctuations in BRD4 concentration, which are inevitable because growth toward macroscopic phase separation locally depletes dilute BRD4 and generates concentration gradients that couple distinct nucleation sites \cite{hensley2022self}.

By contrast, when chromatin tension is sufficiently low to permit co-condensation, this mechanism instead defines a qualitatively different regime for spatiotemporal targeting (\figref{fig:7}b).
In subsaturated conditions ($S<1$), off-chromatin nucleation of the bulk condensed phase is thermodynamically forbidden within this model, so $\joff=0$ for the bulk-nucleation process considered here.
Stable assembly then requires exceeding a threshold acetylation density $\pthr \approx 0.5$, which persists for WT BRD4 over approximately an order of magnitude variation in subsaturated concentration.
This produces strong discrimination between highly acetylated regions and background chromatin with lower but nonzero acetylation.
Because co-condensation in this regime is both thermodynamically stable and kinetically rapid---with a predicted mean assembly time of approximately \SI{90}{ms} at $S=0.1$ for a target region containing 20 acetylated nucleosomes---assemblies can form efficiently in response to changes in acetylation while maintaining strong target--background discrimination over variations in the far-field BRD4 concentration.

Comparison with the single-active-domain \mBD2 model (\figref{fig:7}c) highlights the role of multivalency in achieving sensitivity to acetylation density.
Although \mBD2 also supports co-condensation at low tension, the range of subsaturated conditions that lead to stable and rapid assembly is reduced relative to WT BRD4.
Moreover, the threshold acetylation density varies more strongly with saturation than for WT BRD4, reducing the robustness of discrimination between highly acetylated regions and background chromatin.
Thus, multivalency enhances the ability of co-condensation to distinguish between highly acetylated regions and background chromatin while preserving robustness to variations in BRD4 concentration.

\section{Conclusions}

We have identified a physical regime in which BRD4 assemblies strongly discriminate highly acetylated chromatin from chromatin with lower background acetylation.
Building on prior theoretical studies of co-condensation \cite{Sommer2022,Rouches2025,Renger2022,Quail2021,Shrinivas2019,Sharma2021}, we address the question of whether this mechanism can provide the degree of spatial selectivity required for chromatin targeting without requiring fine-tuned protein concentrations.
Our ultra-coarse-grained molecular-dynamics model demonstrates that the interplay of subsaturation, chromatin mechanics, and multivalent binding defines a parameter regime in which finite BRD4 assemblies form rapidly at highly acetylated regions while suppressing off-target aggregation.
In this regime, strong discrimination between target and background chromatin and robustness to concentration fluctuations emerge simultaneously rather than being in competition.

Our results also help clarify the physical description of chromatin-associated condensates.
The targeting regime characterized here differs from polymer bridging \cite{Nordenskild2024} or collapse driven solely by 1:1 binding interactions~\cite{Brackley2013,Ryu2021}, as it requires that BRD4 possess the intrinsic capacity to phase separate at sufficiently high concentrations through IDR-mediated self-attraction, which is strongest in the C-terminal region.
Nonetheless, this mechanism does not entail bulk phase separation; rather, it defines a regime in which BRD4 assemblies exist as finite, chromatin-associated structures that are stabilized by the combined effects of binding and self-attraction.
This regime enables selective and rapid targeting because assembly occurs under subsaturated conditions with approximately diffusion-limited kinetics rather than being controlled by a large nucleation barrier, while multivalent binding maintains strong discrimination between acetylation levels over a broad range of BRD4 concentrations.

These findings underscore the nuance required when interpreting signatures of phase separation in the context of chromatin \cite{Sabari2018, Chong2018, Woringer2018, Alberti2019, Narlikar2021, Lyons2023, Sabari2024, Bremer2025, Fan2026}.
Features often associated with phase separation---such as thresholded assembly or droplet-like morphology---do not necessarily imply classical nucleation mechanisms or macroscopic demixing.
Conversely, the absence of coarsening or large-scale phase separation does not preclude a thermodynamic origin rooted in intermolecular attractions that drive phase separation at higher concentrations.
Our work further highlights qualitative differences between the formation kinetics of finite chromatin-associated assemblies and the classical nucleation kinetics of phase-separated droplets.

These predicted requirements for selective and concentration-robust targeting can be tested experimentally by varying chromatin acetylation density and global BRD4 concentration while monitoring assembly at defined chromatin regions.
Our simulation results can also be validated by perturbing BRD4 valency to test its predicted role in maintaining discrimination across concentrations.
Although the present model omits additional nuclear components and crowding effects, the core mechanism is not expected to depend qualitatively on these details.
Incorporating such factors will be important for understanding how chromatin-associated BRD4 assemblies recruit downstream transcriptional machinery and operate within the broader nuclear environment.\\

\begin{acknowledgments}
  Research reported in this publication was supported by the National Institute of General Medical Sciences of the National Institutes of Health under award number R35GM155017 to WMJ.
\end{acknowledgments}

\clearpage
\onecolumngrid
\renewcommand\thesection{S\arabic{section}}
\setcounter{section}{0}
\renewcommand\thefigure{S\arabic{figure}}
\setcounter{figure}{0}
\renewcommand\theequation{S\arabic{equation}}    
\setcounter{equation}{0}
\renewcommand\thetable{S\arabic{table}}
\setcounter{table}{0}

\begin{center}
  \textbf{Supplementary Information for ``Co-condensation and multivalency enable acetylation-sensitive, concentration-robust assembly of BRD4 condensates''}
\end{center}

\section{Ultra-coarse-grained (UCG) model of BRD4 and chromatin}
\label{sec:model}

Energies are expressed in units of the thermal energy $k_{\mathrm B}T$, so that $k_{\mathrm B}T=1$ in reduced simulation units.
The background solvent is represented implicitly; energies in the UCG model therefore represent potentials of mean force.
Masses are normalized such that the total mass of a BRD4 molecule is $\mWT=1$.

\subsection{Chromatin model and bonding interactions}

Chromatin is modeled as a semiflexible polymer of nucleosomes connected by linker DNA.
Each nucleosome is represented as a bead with fixed geometry, and neighboring nucleosomes are connected by modified FENE (finite extensible nonlinear elastic)-like bonds that enforce connectivity while allowing finite extensibility \cite{Kremer1990}.

\paragraph{Bending potential.}
A bending potential is applied to successive bonds between nucleosomes, indexed by $i$,
\begin{equation}
\frac{U}{k_{\mathrm B}T}
=
-B \sum_i \cos(\theta_{i+1}-\theta_i),
\end{equation}
where $\theta_i$ is the orientation angle of bond $i$ and $B$ is the dimensionless bending stiffness.
Matching to the continuum worm-like-chain model gives
\begin{equation}
B = \frac{\lp^{\rm chromatin}}{\delta l}.
\end{equation}
Using $\lp^{\rm chromatin} \approx \SI{16.5}{nm}$ and $\delta l \approx \SI{17.8}{nm}$ \cite{Beltran2019,Karymov2001}, we obtain $B \approx 0.93$, corresponding to a bending stiffness of approximately $0.93\,k_{\mathrm B}T$.

We note that chromatin rigidity can, in principle, oppose collapse during co-condensation by introducing a mechanical barrier to bending.
However, sensitivity tests show that this effect is negligible for biologically relevant parameters: only for $\lp^{\rm chromatin} / \delta l \gtrsim 20$ does bending rigidity significantly stabilize extended configurations and impede collapse.
In the parameter regime considered here, chromatin mechanics are therefore well captured by the baseline choice $B \approx 0.93$.

\paragraph{Bonding potential.}
The bonding potential is
\begin{equation}
  \label{eq:Ubond}
\frac{\Ubond(r)}{\kT}  = -\frac{K R_0^2}{2} \ln \left[ 1 - \left(\frac{r - \Delta_{ij}}{R_0}\right)^2 \right] + 4\varepsilon \left[\left(\frac{\sigma_{ij}^*}{r}\right)^{12} - \left(\frac{\sigma_{ij}^*}{r}\right)^6\right],
\end{equation}
where $\Delta_{ij} = \sigma_{ij}^* = (\sigma_i^* + \sigma_j^*)/2$ offsets the effective attachment points between beads $i$ and $j$ to account for finite bead size, $\sigma_i^*$ denotes the hard-core diameter of bead $i$, and $\varepsilon=1$ sets the repulsive energy scale.
This form is a modified FENE bond.
Whereas the standard FENE potential \cite{Kremer1990} assumes that bonds connect particle centers, bead diameters are comparable to bond lengths in the present model, so that a center-to-center connection would lead to unphysical overlap and incorrect geometry.
To account for the bead size, the effective attachment points of the bond are shifted by $\Delta_{ij}$, which can be interpreted as moving the endpoints of the spring from bead centers toward their surfaces.
With this modification, the maximal bond extension becomes $R_0 + \Delta_{ij}$, ensuring physically meaningful separation between bonded beads.
The same bonding potential is used consistently for both chromatin linkers and protein domains connected by intrinsically disordered regions (IDRs), providing a unified treatment of connectivity across the model.

\begin{figure*}[h]
    \centering
    \includegraphics[width=0.5\textwidth]{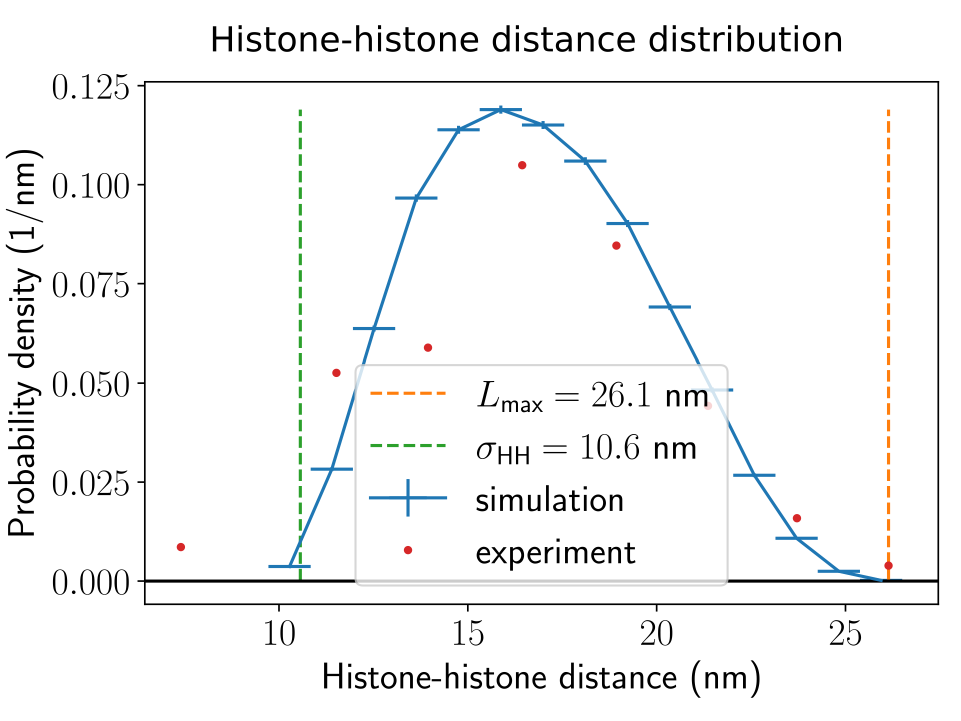}
    \caption{Parametrization of the rigidity of the nucleosome--nucleosome bond using experimentally measured distance fluctuations. Red data points are reproduced from Ref.~\cite{Karymov2001}. The blue line shows statistics from chromatin-only simulations at low tension ($\FT \approx 0.3$ pN). The green dashed line indicates the WCA nucleosome size. Because the geometry of a real nucleosome is not spherical but disk-like, there are orientations in which two nucleosomes can approach more closely than $\sgmHH$. Some experimental distances therefore fall below the model value of $\sgmHH$, since the UCG model approximates each nucleosome as an equivalent-volume sphere. The orange dashed line shows the maximum extension between nucleosome centers, given by $\Llinker + \sgmHH$.}
    \label{fig:1S2}
\end{figure*}

Chromatin linker lengths are set by the number of DNA base pairs in the linker, $R_0 = L_{\mathrm{linker}} = N_{\mathrm{bp}} l_{\mathrm{bp}}$, where $l_{\mathrm{bp}} = \SI{0.34}{nm}$ \cite{Cheng2012}.
For U2OS cells, which we take as a representative system, we use $N_{\mathrm{bp}} \sim 40$, giving $R_0 \approx \SI{14}{nm}$.
Chromatin stiffness is parameterized using a persistence length $\lp^{\rm chromatin} = \SI{16.5}{nm}$ \cite{Beltran2019}.
We estimate the chromatin rigidity parameter $K$ from the persistence length using $K = 3/(2 \lp^{\rm chromatin} L_{\mathrm{linker}})$, which relates the spring constant for small end-to-end distance fluctuations to the worm-like-chain variance in the short-linker limit.
The chromatin rigidity parameter $K$, initially estimated as $6.4 \times 10^{-3}\,\mathrm{nm}^{-2}$, is scaled down by a factor of approximately $0.7$ to better match the experimental nucleosome--nucleosome distance distribution (\figref{fig:1S2}).
To conserve the steric volume of a nucleosome disk of thickness $\SI{6.5}{nm}$ and diameter $\SI{11}{nm}$ \cite{Richmond1984}, we use an effective nucleosome diameter $\sigma^* = \SI{10.6}{nm}$.

\paragraph{Nucleosome geometry and acetylation.}
Each nucleosome contains eight histone tails in total, of which two H3 and two H4 tails are acetylatable.
To represent their spatial organization, we place tail attachment points approximately at the vertices of a cube surrounding the nucleosome core \cite{Richmond1984}.
This geometry captures the roughly isotropic distribution of tails while preventing artificial clustering.
To maintain this arrangement, we introduce 12 weak angular constraints between tail--nucleosome--tail triplets corresponding to the edges of the cube.
These constraints are chosen to be minimal, serving only to prevent collapse of tails toward a single side of the nucleosome (e.g., toward a nearby BRD4-rich region) while preserving sufficient flexibility for realistic binding configurations.

Acetylation (Ac) is introduced on a segment of $\Nac$ consecutive nucleosomes.
Each H3 and H4 tail is acetylated independently with probability $\pAc$ unless otherwise specified.
To connect $\Nac$ to experimentally observed chromatin organization, we estimate the typical number of consecutively acetylated nucleosomes from genomic data.
Acetylated regions in human cells span a wide range of lengths, from $\sim \SI{100}{bp}$ in small enhancers to $>\SI{50}{kbp}$ in super-enhancers \cite{Panigrahi2021, Wu2023}, with a substantial fraction of peaks in the range $\sim 1$--$\SI{6}{kbp}$ \cite{Wu2023}.
A nucleosome wraps $\sim \SI{150}{bp}$ of DNA, and linker DNA typically contributes an additional $\sim 10$--$\SI{80}{bp}$ \cite{Wu2023}.
For U2OS cells, the linker length of $\sim \SI{40}{bp}$ gives $\sim \SI{190}{bp}$ per nucleosome.
This correspondence implies that acetylated regions of $\sim 1$--$\SI{6}{kbp}$ span approximately $5$--$30$ consecutive nucleosomes.
Guided by this estimate, we use $\Nac = 20$ in the main text and also consider $\Nac = 30$ in the supplementary figures.

To limit Ac density fluctuations, we use a ``shuffle'' acetylation scheme, in which a fixed number $4\pAc\Nac$ of acetylated tails are randomly distributed among the $4\Nac$ sites. This removes fluctuations in the overall acetylation density while preserving local randomness. This scheme is motivated by the strong sensitivity of observables near the acetylation threshold for co-condensation $\pAc \approx \pthr$, where fluctuations in the total number of acetylated tails significantly affect quantities such as the average assembly size.

\subsection{BRD4 model and IDR-mediated interactions}

\paragraph{Molecular geometry and excluded volume.}
Each BRD4 molecule is represented by three beads corresponding to BD1, BD2, and the C-terminal domain.
This representation is motivated by the domain architecture of BRD4, which contains two bromodomains (BD1 and BD2) responsible for binding acetylated histone tails, and a C-terminal structured domain embedded within extended IDRs \cite{Strom2024}.
Domain boundaries and structured regions are identified from AlphaFold2 structural predictions \cite{Jumper2021}, using high-confidence segments to define globular domains and assigning intervening regions to IDRs.
Effective sizes $\sigma_i^*$ of globular domains are determined from these structural models by matching their volumes.
For each domain, atoms corresponding to the selected amino-acid range are used to construct an inertia tensor, which is then mapped to an equivalent uniform ellipsoid.
The resulting effective diameter $\sigma_i^*$ is defined such that the ellipsoid volume is preserved, providing a consistent measure of steric size across domains.
Steric parameters are chosen not only to prevent overlap but also to reproduce physically reasonable condensed-phase densities.
Accordingly, effective bead sizes incorporate both globular domain volume and associated IDR volume, ensuring that excluded-volume interactions yield realistic packing in the dense phase.

To assign IDR segments to individual coarse-grained beads, we partition the full BRD4 amino-acid sequence based on the locations of structured domains.
IDR regions are defined as the sequence segments between adjacent structured domains, with boundaries assigned at the midpoints between neighboring domain endpoints.
Concretely, if structured domains have index ranges $[i^*_{\mathrm{beg}}, i^*_{\mathrm{end}})$, then the IDR associated with a given domain consists of all amino acids between the midpoint to the preceding domain and the midpoint to the following domain, excluding residues belonging to structured regions.
This ensures that each amino acid in the sequence is uniquely assigned either to a globular domain or to a single IDR segment.
This construction provides an unambiguous and systematic partitioning of the sequence, allowing the IDR length $\nIDR$ associated with each coarse-grained bead to be defined consistently from the underlying protein sequence.

To account for the contribution of IDRs, we construct their effective volume from amino-acid volumes,
\begin{equation}
V_{\mathrm{IDR},i} = \frac{1}{0.7} \sum_{k \in \mathrm{IDR}_i} v_{\mathrm{AA},k},
\end{equation}
where $v_{\mathrm{AA},k}$ are residue-specific volumes \cite{Zamyatnin1984} in solution.
The prefactor $1/0.7$ accounts for hydration and imperfect packing, reflecting that IDRs in solution do not fill space at unit density but instead occupy an expanded, solvent-accessible volume.
This construction corresponds to an approximate picture in which IDR segments compactify around structured domains in the dense phase while retaining a finite excluded volume set by their sequence composition.

Excluded-volume interactions are modeled using a Weeks--Chandler--Andersen (WCA) potential,
\begin{equation}
  \label{eq:UWCA}
\frac{\Uwca}{\kT} = 4\varepsilon \left[\left(\frac{\sigma_{ij}}{r}\right)^{12} - \left(\frac{\sigma_{ij}}{r}\right)^6\right] + U_0, \quad r \le \Rcut,
\end{equation}
with $\sigma_{ij} = (\sigma_i + \sigma_j)/2$ and $\varepsilon = 1$.
The offset $U_0$ is chosen to make $\Uwca(\Rcut = 2^{1/6} \sigma_{ij})=0$.
Effective bead sizes $\sigma_i$ are determined from their globular domain size $\sigma_i^*$ and IDR volume,
\begin{equation}
\sigma_i = \left((\sigma_i^*)^3 + \frac{6 V_{\mathrm{IDR},i}}{\pi}\right)^{1/3}.
\end{equation}

\paragraph{Bonding potential.}
Bonded interactions between the BRD4 beads follow \eqref{eq:Ubond}.
For protein IDRs, the maximum extension is set by the contour length, $R_0 = \lIDR = \nIDR \lAA$, where $\lAA = \SI{0.38}{nm}$ is the average spacing between amino acids \cite{Sarkar2005, Erickson1994, Trombits1998}.
The bond stiffness $K$ is chosen to reproduce small extension fluctuations of polymer segments.
For IDR-mediated connections, this gives $K = 3 / (2 \lp^{\rm IDR} \lIDR)$, where the IDR persistence length is $\lp^{\rm IDR} \approx \SI{4}{\AA}$.

\paragraph{IDR-mediated attractive interactions.}
Non-specific attraction between IDR-containing beads is modeled using a Gaussian potential,
\begin{equation}
  \label{eq:Ugauss}
\frac{\Ugauss}{\kT} = -A \lAA^3 \frac{\nIDR[i] \nIDR[j]}{(\Rgi^{2} + \Rgj^2)^{3/2}}  \exp\left[ - \frac{1.6 r^2}{\Rgi^{2} + \Rgj^2} \right],
\end{equation}
where $\Rgi^2 = (\sigma_i^*/2)^2 + \lp^{\rm IDR} \lAA \nIDR[i] / 3$.
This interaction form is motivated by Flory-type polymer theory.
The original derivation assumes that polymer chains adopt approximately Gaussian density profiles around their centers of mass \cite{Flory1949}, which leads to an effective interaction between chains proportional to the overlap of these densities.
This framework was later generalized to interactions between heterogeneous chains of different sizes and compositions \cite{Flory1950}, yielding the form used here.
The Gaussian factor therefore reflects the physical picture that IDRs behave as diffuse polymer blobs whose overlap generates an effective attraction.
The numerical factor in the exponential, here taken as $1.6$, arises from more detailed simulations of polymer chains beyond the idealized Gaussian model, which show that the effective interaction range is slightly shorter than predicted by the simplest theory \cite{Louis2000}.

The prefactor $A$ encodes the strength of the IDR-mediated interaction.
This overall interaction strength is calibrated using corelet phase-separation experiments \cite{Bracha2018, Strom2024}.
Matching the observed critical valence of BRD4 corelets in those experiments yields $A \approx 0.105$ (\figref{fig:1S5}).

Key BRD4 parameters are summarized in \tabref{tbl:BRD4params}.

\begin{table}[h!]
\begin{tabular}{|l|l|l|l|l|l|l|l|l|l|}
\hline
name & $i_{\text{beg}}$ & $i_{\text{end}}$ & $i^*_{\text{beg}}$ & $i^*_{\text{end}}$ & $\sigma^*$ / nm & $\nIDR$ & $\sigma$ / nm & mass / $m_{\text{BRD4}}$ & $2 \Rg$ / nm \\ \hline
BD1  & {[}1      & 259)      & {[}61       & 164)        & 3.36                & 155       & 4.48       & 0.188                 & 6.54           \\ \hline
BD2  & {[}259    & 528)      & {[}353      & 456)        & 3.35                & 166       & 4.55       & 0.196                 & 6.70           \\ \hline
C    & {[}528    & 1362{]}   & {[}600      & 679)        & 3.26                & 756       & 6.75              & 0.616                 & 12.80          \\ \hline
\end{tabular}
\caption{Parameters of the BRD4 coarse-grained model, including sequence index ranges for each domain, structured-domain boundaries used for size estimation, and derived coarse-grained properties.}
\label{tbl:BRD4params}
\end{table}

\subsection{BRD4--chromatin interactions}

\paragraph{Binding strengths and dissociation constants.}
Binding between BRD4 and chromatin is modeled as 1:1 interactions between bromodomains and acetylated histone tails.
Binding strengths are derived from experimental dissociation constants \cite{Vollmuth2009} using
\begin{equation}
\frac{\Ubind}{\kT} = -\ln\left(v_0 \Kd\right),
\end{equation}
where $v_0$ is the interaction volume and $\Kd$ is the dissociation constant.
This expression can be understood from a balance between entropic and enthalpic contributions to binding.
When a bromodomain binds an acetylated lysine, the molecule is effectively confined from a dilute volume $1/c$ to a binding volume $v_0$, incurring an entropic penalty $\sim -\ln(c v_0)$.
At equilibrium, this penalty is compensated by a free energy $\Ubind$, which by definition balances at $c = \Kd$, yielding $\Ubind/\kT = -\ln(v_0 \Kd)$.
In applying this relation, we assume that bromodomains measured in isolation in titration experiments have approximately the same effective size and binding geometry as bromodomains within the full BRD4 molecule.
We therefore estimate the interaction volume as $v_0 \sim V_{\mathrm{BD}} = (\pi/6)(\sigma_i^*)^3$, corresponding to the volume of the coarse-grained globular domain.

\begin{figure*}[h]
    \centering
    \includegraphics[width=1\textwidth]{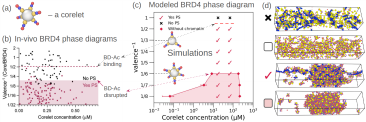}
    \caption{Parametrization of two interaction strength values of the UCG model based on two \textit{in vivo} experimental phase diagrams.
(a)~A UCG model of a 6-valent corelet~\cite{Strom2024} used to fit IDR--IDR and BD--Ac interaction strengths.
(b)~\textit{In vivo} experimental phase-diagram boundaries for systems with wild-type binding between acetylated (Ac) tails and BRD4 bromodomains (BDs) (dashed line, critical corelet valence = 2), and with disrupted Ac--BD binding (shaded region, critical valence = 6); data are reproduced from Ref.~\cite{Strom2024}.
(c)~\textit{In silico} phase diagrams for the UCG model corresponding to (b). Insets on the left show corelet systems with corelet valence 3 and 6. The shaded region indicates phase separation without chromatin. The IDR--IDR attraction strength is tuned to yield a critical corelet valence of 6. Checks/crosses denote the presence/absence of phase separation with fully acetylated chromatin. The BD--Ac binding strength is tuned to be the lowest value that reliably produces phase separation at corelet valence = 2.
(d)~Illustrations of representative cases from (c) are labeled as follows.
Cross: example with no phase separation in the presence of chromatin.
Open square: example with no phase separation in the absence of chromatin.
Check: example with phase separation in the presence of fully acetylated chromatin.
Filled square: example with phase separation in the absence of chromatin.}
    \label{fig:1S5}
\end{figure*}

Experimental measurements report four distinct dissociation constants for BD1/BD2 binding to H3/H4 tails \cite{Vollmuth2009}.
These yield slightly different binding energies $U_{\text{BD1-H3}} : U_{\text{BD1-H4}} : U_{\text{BD2-H3}} : U_{\text{BD2-H4}} = 6.57 : 5.56 : 5.55 : 6.58$.
Accounting for experimental uncertainty, we group them into two effective interaction strengths with the ratio $6.57 : 5.55$ corresponding to stronger and weaker binding pairs, preserving the overall scale while simplifying the model.
Importantly, these values correspond to binding of a single acetylated lysine, whereas histone tails can contain multiple acetylation sites that contribute cooperatively to binding \cite{Dey2003}.
This motivates introducing an overall scaling factor to account for multivalent contributions along a single histone tail.
We calibrate this factor by matching phase-separation behavior in corelet systems \cite{Strom2024}, choosing the minimal scaling that reliably produces phase separation at valence $v=2$ (\figref{fig:1S5}).
Calibration yields a multiplicative factor of $2.15$; this is consistent with the fact that the typical number of acetylations per histone tail is two, with some tails having more, yielding an expected enhancement in the effective binding strength by a factor slightly greater than two.
This procedure ensures that relative binding affinities are preserved while the absolute scale is tuned to reproduce experimentally observed phase behavior.

\paragraph{Enforcement of 1:1 binding interactions.}
To enforce strict 1:1 binding while maintaining numerical stability and minimal perturbation to the rest of the model, we implement the following modifications to the nonbonded interactions, \eqref{eq:UWCA} and \eqref{eq:Ugauss}, with explicit geometric and dynamical constraints:
\begin{enumerate}
\item Disable steric WCA repulsion between BD domains and acetylated tails, allowing an Ac tail to occupy the center of the binding pocket.
\item Introduce a Gaussian attraction with well depth $U_{\text{bind}}$ and a short interaction range. The range is chosen as $R_{\mathrm{bind}} \approx 0.34 \sigma_{\mathrm{BDAc}}/2$, where $\sigma_{\mathrm{BDAc}} = (\sigma_{\mathrm{BD}} + \sigma_{\mathrm{Ac}})/2$, and $\sigma_{\mathrm{BD}}$ and $\sigma_{\mathrm{Ac}}$ are the WCA diameters of a BD bead and an acetylated-tail bead, respectively. This choice ensures that the attraction is significantly shorter-ranged than the steric exclusion between BD domains. It therefore enforces geometric exclusivity: once an Ac tail binds and localizes near the center of a BD domain, other BD domains cannot approach closely enough to feel the binding potential. Steric repulsion limits the closest approach of two BD domains to $r \gtrsim \sigma_{\mathrm{BDBD}}$, where $\sigma_{\mathrm{BDBD}}$ is the BD--BD WCA contact diameter; at this separation, $|U_{\text{bind}}(r)| \ll \kT$.
\item Introduce steric repulsion between acetylated tails with effective size $\sigma_{\mathrm{AcAc}} \approx 0.85 \sigma_{\mathrm{BDBD}}$. This prevents multiple Ac tails from simultaneously approaching a bound BD domain. The value is chosen as a compromise, as it is large enough to enforce single occupancy, but small enough to avoid artificially disrupting droplet structure or histone tail packing.
\item Extend the maximum bond length of histone tails from $\lIDR/2$ to $3\lIDR/4$, where $\lIDR$ is the length of the IDR portion of the corresponding histone tail. Short linker lengths combined with nonlinear bond potentials would otherwise generate large forces over short distances, requiring prohibitively small integration timesteps. This extension reduces stiffness without altering the qualitative geometry of binding.
\item Increase the mass of acetylated tails to $m_{\mathrm{Ac}} \approx 0.6\, m_{\mathrm{BRD4}}$. In the coarse-grained representation, tails would otherwise have very small mass, leading to large accelerations under strong, short-range binding forces. Increasing the mass stabilizes integration while primarily affecting dynamics rather than equilibrium properties, making it the least disruptive modification.
\end{enumerate}
These combined modifications ensure strict 1:1 binding.
Validation simulations comprising $\sim 1000$ BRD4 molecules and $\sim 800$ acetylated tails showed no violations of the 1:1 binding constraint.

Together, these elements define a minimal ultra-coarse-grained model capturing multivalent binding, self-attraction, and chromatin mechanics required for BRD4 assembly.

\section{Simulation methods}
\label{sec:simulations}

All simulations are implemented in LAMMPS~\cite{Thompson2022}.
Different ensembles are utilized for distinct simulation purposes, as described below.
In all simulations, we use a Langevin thermostat \cite{Ermak1980} with damping time constant $\tauLangevin=10$ in reduced simulation units.
This value is chosen as a compromise between maintaining sufficiently rapid diffusion for efficient sampling and limiting the concentration dependence of the BRD4 diffusion coefficient $D(S)$ arising from intermolecular collisions.

\subsection{Calibration of dynamical timescales}
\label{sec:time_calibration}

To map the simulated dynamics to physical timescales, we use the experimentally measured intranuclear diffusion coefficient of freely diffusing BRD4, $D_{\rm exp} = \SI{5.4 \pm 0.3}{\mu m^2/s}$ \cite{Xie2022}.
As an independent order-of-magnitude consistency check, the measured intranuclear viscosity, $\eta \approx \SI{4\pm1}{cP}$~\cite{Siebrasse2008}, together with the Stokes--Einstein relation~\cite{Sutherland1905,Einstein1905} and an effective hydrodynamic radius $R_{\rm h} \sim R_{\rm g} \approx \SI{6}{nm}$, gives $D \approx \SI{9.3\pm2.4}{\mu m^2/s}$.
Using the direct experimental measurement of $D_{\rm exp}$ for the quantitative calibration, we obtain a characteristic timescale for a BRD4 molecule to diffuse one molecular diameter,
\begin{equation}
    t_{\rm Diff}^{\rm exp}
    =
    \frac{(2R_{\rm g})^2}{6D_{\rm exp}}
    \approx
    \SI{4.4\pm0.3}{\mu s}.
    \label{eq:tdiff_exp}
\end{equation}
We use this calibration to convert the assembly timescales measured in simulations to physical units; given the experimental and modeling uncertainties, we consider these timescales to be order-of-magnitude estimates.
This procedure avoids assigning physical significance to the bare molecular-dynamics time unit and instead calibrates the kinetic observables directly to the experimentally measured BRD4 mobility.

\subsection{$NVT$ simulations of bulk coexistence}

Bulk phase behavior is characterized using constant-volume ($NVT$) simulations.
Systems are initialized with uniform BRD4 concentration and evolved until phase separation occurs, yielding coexisting dilute and condensed phases.
The dilute-phase concentration $\ccoex$ is determined from the equilibrium density of the dilute region in the absence of chromatin.

In simulations with chromatin, the chromatin fiber spans the simulation box under periodic boundary conditions along the $z$ direction, ensuring coexistence between phases in direct contact with the polymer.
Concentration profiles are computed by averaging particle concentrations along the $z$ axis after equilibration.

\subsection{Determination of the coexistence concentration $\ccoex$}

The coexistence concentration $\ccoex$ is determined from $NVT$ simulations of BRD4 in the absence of chromatin.
Systems are initialized at concentrations above coexistence and allowed to phase separate into coexisting dilute and condensed phases.
The dilute-phase concentration is measured by averaging the BRD4 density in regions far from the condensed phase after equilibration.
To reduce finite-size effects, measurements are performed over multiple independent simulations and averaged over time once equilibrium coexistence is established.
The resulting value of $\ccoex$ is used to define the saturation ratio $S = c_{\rm far} / \ccoex$ throughout the main text.

\subsection{$NP_xP_yL_zT$ ensemble and chromatin tension control}
\label{sec:npxpylzt}

To control the far-field BRD4 concentration and chromatin mechanics independently, we perform simulations in an $NP_xP_yL_zT$ ensemble.
We use this ensemble for both equilibrium simulations of chromatin-associated assemblies at subsaturated conditions ($S<1$) and kinetic measurements of assembly formation and condensed-phase nucleation rates below ($S<1$) and above ($S>1$) coexistence, respectively.
This simulation approach suppresses finite-simulation-volume artifacts due to BRD4 depletion while still producing continuous trajectories with Langevin dynamics, as described below.

In this ensemble, the number of particles $N$, temperature $T$, and box length $L_z$ are fixed, while the lateral box dimensions $L_x$ and $L_y$ fluctuate under control of the corresponding pressures $P_x$ and $P_y$.
Periodic boundary conditions are applied in all three directions.
Because the chromatin fiber spans the simulation box along the $z$ direction, $L_z$ is excluded from the barostat so that the mechanical constraint imposed on the chromatin is not coupled directly to the pressure used to control the far-field BRD4 concentration.

In the simulation conditions considered in this work, both the far-field BRD4 concentration at subsaturated conditions ($S<1$) and the initially homogeneous BRD4 concentration at supersaturated conditions prior to nucleation ($S>1$) are well approximated by the ideal equation of state,
\begin{equation}
    P_{\rm gas} \simeq c_{\rm far} k_{\mathrm B}T.
\end{equation}
We therefore impose the lateral pressures
\begin{equation}
    P_x=P_y=S\ccoex k_{\mathrm B}T,
\end{equation}
where $\ccoex$ is the dilute-phase BRD4 concentration at bulk coexistence obtained independently from $NVT$ direct-coexistence simulations.
At subsaturated conditions ($S<1$), the barostat enforces a far-field concentration
\begin{equation}
    c_{\rm far} \simeq S\ccoex
\end{equation}
throughout the simulation.
For nucleation calculations at $S>1$, the initially homogeneous system has $c=c_{\rm far}\simeq S\ccoex$ prior to nucleation, while pressure control maintains the far-field concentration during the early stages of nucleation and post-critical growth; we terminate such simulations well before the condensed phase occupies the entire simulation volume.

Chromatin extension is controlled by the fixed box length $L_z$.
We define
\begin{equation}
    \lambda = \frac{\Nnucleosomes \aHH}{L_z},
\end{equation}
where $\aHH$ is the equilibrium spacing between neighboring nucleosomes.
Large $\lambda$ corresponds to low initial chromatin tension and permits substantial local compaction, whereas $\lambda\approx1$ corresponds to a nearly straight, high-tension fiber.
The mapping between $\lambda$ and the resulting mechanical tension $\FT$ is obtained from independent simulations without BRD4 (\figref{fig:3S1}).
Although $P_z$ is not externally controlled, the far-field BRD4 gas equilibrates to an isotropic pressure tensor; however, because tension on the chromatin fiber also contributes to the $z$ component of the total stress tensor, the measured $P_z$ of the simulation box may not be exactly equal to $P_x$ and $P_y$ at equilibrium.

\begin{figure*}
    \centering
    \includegraphics[width=1\textwidth]{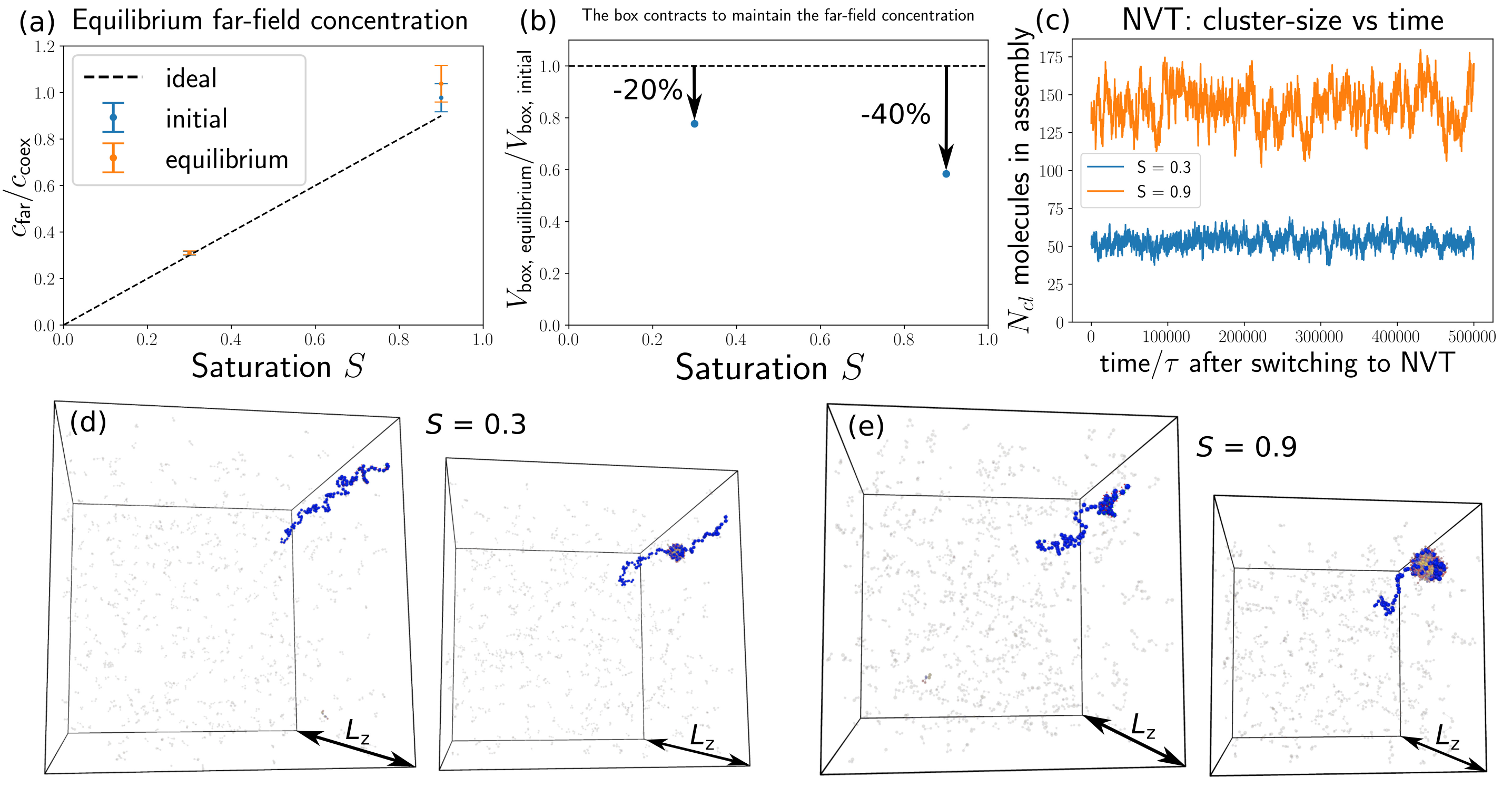}
    \caption{Validation of far-field concentration control in the $NP_xP_yL_zT$ ensemble.
      (a)~Far-field BRD4 concentration measured at two representative subsaturated conditions, excluding the region surrounding the chromatin and chromatin-associated assembly.  The far-field concentration remains approximately constant at the target value $c_{\rm far}\approx S\ccoex$ before and after formation of the dense assembly.
      (b)~The corresponding simulation-box volume. Assembly formation produces a substantial change in the box volume as the lateral box dimensions respond to sequestration of BRD4 into the chromatin-associated assembly, while the far-field concentration in (a) remains approximately constant.
      (c)~Further validation by switching equilibrated configurations from the $NP_xP_yL_zT$ ensemble to the $NVT$ ensemble at the equilibrated box dimensions.  No appreciable change in the chromatin-associated assembly or surrounding dilute phase is observed after the switch, demonstrating that the equilibrated finite assembly is not stabilized by continued pressure control.  Together, these results confirm that the $NP_xP_yL_zT$ simulations maintain the prescribed far-field BRD4 concentration while allowing the simulation volume to adjust during finite chromatin-associated self-assembly.
      Representative snapshots of the system before (left) and after (right) assembly formation are shown for the simulations at (d) $S=0.3$ and (e) $S=0.9$; in each panel, the relative box sizes before and after assembly formation are shown to scale, illustrating the substantial change in system volume.}
    \label{fig:npxpy_validation}
\end{figure*}

We directly validate the far-field concentration control in \figref{fig:npxpy_validation}.
For two representative values of $S<1$, the BRD4 concentration measured sufficiently far from the chromatin and excluding the region surrounding the chromatin-associated assembly remains close to the target value $S\ccoex$ both before and after assembly formation.
This is accompanied by a decrease in the simulation-box volume of approximately $20\%$ at $S=0.3$ and $40\%$ at $S=0.9$, as the lateral dimensions respond to sequestration of BRD4 into the finite assembly.
The substantial volume response demonstrates that the constancy of $c_{\rm far}$ is maintained by the pressure-controlled ensemble rather than arising simply because assembly formation produces a negligible perturbation.
As a further test, after the chromatin-associated assembly has equilibrated in the $NP_xP_yL_zT$ ensemble, we fix the simulation box at its equilibrated dimensions and continue the same configuration in the $NVT$ ensemble.
The assembly and surrounding BRD4 gas show no appreciable change following this switch (\figref{fig:npxpy_validation}), demonstrating that the equilibrated finite assembly is not stabilized artificially by continued pressure control.

\subsection{Finite assembly stability and hysteresis}

Assemblies are identified using a cluster-detection algorithm based on spatial proximity.
Two BRD4 molecules are considered part of the same assembly if any pair of their constituent beads lies within a specified cutoff distance corresponding to the range of attractive interactions.
The assembly size $\Ncl$ is defined as the number of BRD4 molecules in the largest connected cluster. This definition is used consistently across equilibrium and kinetic analyses.
Radial concentration profiles for finite assemblies are computed by averaging particle densities as a function of distance from the assembly center of mass.
For bulk coexistence simulations, concentration profiles are computed along the $z$ direction.
These measurements allow direct comparison between finite assemblies formed under subsaturated conditions and bulk condensed phases in direct-coexistence simulations.

The stability of finite BRD4 assemblies under subsaturated conditions is characterized by comparing simulations initialized with and without a pre-formed BRD4 cluster.
In simulations initialized from a dispersed state, assembly formation is monitored to determine whether stable co-condensation occurs.
In complementary simulations initialized with a pre-formed cluster, the persistence or dissolution of the assembly is used to determine the stability boundary.
The separation between formation and dissolution conditions defines a hysteresis region associated with a finite free-energy barrier between the dispersed and assembled states.

\subsection{Adaptive biasing force calculations of binding free-energy profiles}

Adaptive biasing force (ABF) calculations are used to obtain BRD4--chromatin binding free-energy profiles under high-tension conditions, where chromatin deformation and co-condensation are suppressed.
The specific ABF setup used to extract effective dissociation constants is described in \secref{sec:multivalency}.
In these calculations, the collective variable is the center-of-mass distance between BRD4 bromodomains and acetylated tails, as defined in \eqref{eq:ABF_CV}.

\section{Multivalency and combinatorial binding theory}
\label{sec:multivalency}

\subsection{Effective binding affinity and $\Kd$ measurements}

\paragraph{Inference via adsorption simulations.}
To quantify BRD4--chromatin interactions in the coarse-grained model, we relate simulation binding parameters to experimentally measured dissociation constants $\Kd$ \cite{Vollmuth2009}.
To validate this mapping in the model, we perform simulations of BRD4 binding to chromatin under conditions where co-condensation is suppressed (high chromatin tension).
Under these conditions, the measured occupancy of BRD4 on chromatin can be related directly to an effective $\Kd$.
The effective dissociation constant is obtained by fitting the fraction of bound BRD4 molecules as a function of concentration to a standard adsorption isotherm.
This procedure allows direct comparison between the wild-type (WT) BRD4 model and the single-active-domain \mBD2 variant, ensuring that both models have approximately matched effective binding affinities at intermediate acetylation density ($\pAc \approx 0.5$), as described in the main text.

\paragraph{Direct calculation via ABF simulations.}
To obtain $\Kd$ directly from simulations, we perform explicit binding free-energy calculations using ABF simulations.
The system consists of a single nucleosome at high tension ($\lambda=1$) with controlled acetylation patterns and a single BRD4 molecule.
Acetylation is assigned independently to H3 and H4 tails, allowing configurations corresponding to $\pAc^{\mathrm{H3}}, \pAc^{\mathrm{H4}} \in \{0, 0.5, 1\}$.
This setup isolates BRD4--chromatin binding from co-condensation effects.

We define the ABF collective variable as the distance between the center of mass of the BRD4 bromodomains and the center of mass of the acetylated tails,
\begin{equation}
\label{eq:ABF_CV}
r =
\left|
\mathbf{R}_{\mathrm{CoM}}(\mathrm{BD1},\mathrm{BD2})
-
\mathbf{R}_{\mathrm{CoM}}(\mathrm{Ac\ tails})
\right|,
\end{equation}
where $\mathbf{R}_{\mathrm{CoM}}(\mathrm{Ac\ tails})$ is computed over the acetylated H3 and H4 tails present in the simulation.
We perform ABF sampling along this coordinate to obtain the free-energy profile $F(r)$.
A configuration is defined as \emph{bound} if the minimum pairwise distance between any bromodomain and any acetylated tail satisfies
\begin{equation}
\rmin = \min_{i \in \mathrm{Ac},\, j \in \{\mathrm{BD1,BD2}\}} r_{ij} < \Rtop,
\end{equation}
where $\Rtop$ is chosen to include the BD--Ac binding well while excluding configurations with more than one bromodomain participating in the interaction.
The total binding probability in a system of finite size can be written as
\begin{equation}
\Pb = \int_0^{L} P(b \mid r)\, p(r)\, dr,
\end{equation}
where $P(b \mid r)$ is the conditional probability of being bound at fixed $r$ and $p(r)$ is the equilibrium radial probability density.
To properly normalize $p(r)$, we define a modified free energy
\begin{equation}
\tilde{F}^{3D}(r) = \frac{F(r)}{\kT} + \ln\left[\left(\frac{r}{\sigma}\right)^2\right],
\end{equation}
where $\sigma$ is an arbitrary reference length that cancels from normalized probabilities.
This removes the entropic contribution from the radial measure.
This allows us to approximate $\tilde{F}^{3D}(r)$ as constant beyond a distance $\Rfar$ where interactions vanish.

Using this construction, the binding probability in a sphere of radius $L$ becomes
\begin{equation}
\label{eq:Pb_expand}
\Pb(L) = \frac{\int_0^{\Rb} e^{-\tilde{F}^{3D}(r)} r^2 P(b|r)\, dr}{\int_0^{\Rfar} e^{-\tilde{F}^{3D}(r)} r^2 dr + \frac{L^3 - \Rfar^3}{3} e^{-\tilde{F}^{3D}(\Rfar)}},
\end{equation}
where $\Rb$ is the largest distance at which $P(b|r)$ is nonzero, and $\Rfar$ is chosen in the noninteracting regime where $\tilde{F}^{3D}(r)$ is approximately constant.
We define the dissociation constant $\Kd$ as the concentration at which $\Pb = 1/2$. Equivalently, if $\Ld$ is the radius of the sphere whose volume gives half binding probability, then
\begin{equation}
\label{eq:Kd_ABF}
\frac{1}{\Kd} = \frac{4}{3}\pi \Ld^3.
\end{equation}

\subsection{Parametrization of the single-active-domain \mBD2 model}

The \mBD2 variant is constructed to isolate the effect of reducing the number of active Ac-binding bromodomains while preserving the steric architecture of BRD4.
Both BD beads are retained in the coarse-grained model, but BD2--Ac binding is disabled, so $\nd=1$ denotes one active Ac-binding domain rather than one physical BD bead.

\begin{figure*}
    \centering
    \includegraphics[width=1\textwidth]{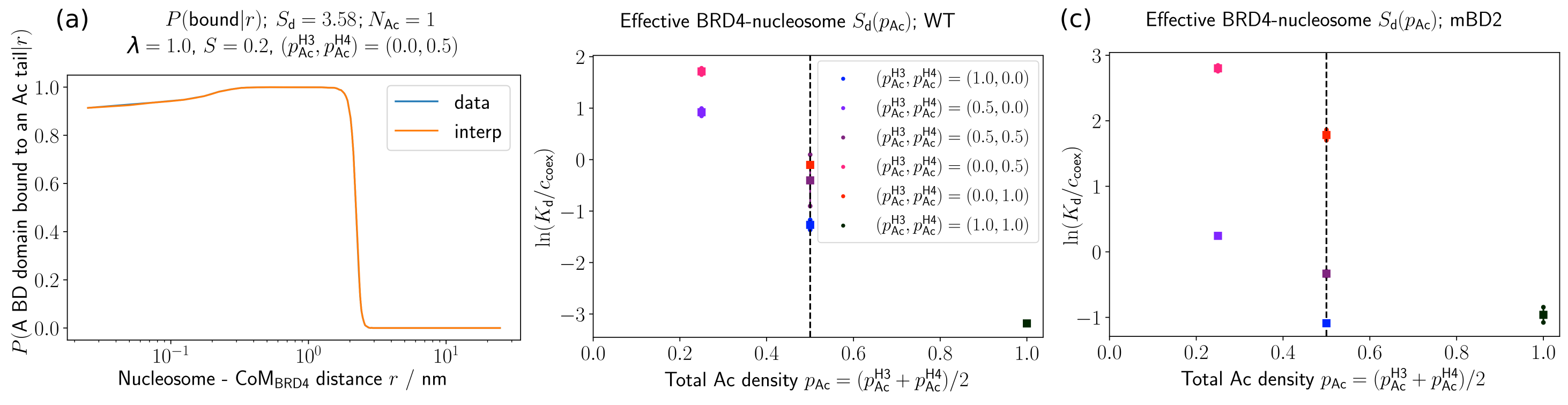}
    \caption{$\Kd$ measurement and \mBD2 construction.
(a)~Representative $P(b \mid r)$ profile used in \eqref{eq:Pb_expand}, shown for $S=0.2$ with one of two H4 tails acetylated. Lines show the measured profile and interpolation used for integration; error bars are within the line thickness.
(b)~Effective $\Kd(\pAc)$ for WT BRD4 measured using \eqref{eq:Kd_ABF}. Different point colors denote different H3/H4 acetylation patterns with the same total acetylation density, $\pAc=(\pAc^{\mathrm{H3}}+\pAc^{\mathrm{H4}})/2$.
(c)~Effective $\Kd(\pAc)$ for the single-active-domain \mBD2 model. The BD1--Ac well depth is increased by a factor of 1.131 relative to WT, while BD2--Ac binding is disabled. This approximately matches the WT effective affinity at $\pAc=0.5$: $\ln(\Smy[d,WT])\approx -0.57$ and $\ln(\Smy[d,\mBD2])\approx -0.39$, where $\Smy[d]\equiv\Kd/\ccoex$, corresponding to agreement within $\sim\pm0.2\,\kT$. At $\pAc=0.5$, the H3-only, H4-only, and mixed H3/H4 acetylation patterns have different combinatorial multiplicities; in particular, the mixed case has $\binom{4}{2}-2\binom{2}{2}=4$ configurations, whereas each H3-only or H4-only case has $\binom{2}{2}=1$ configuration.}
    \label{fig:4S3}
\end{figure*}

To approximately match the effective BRD4--histone tail binding affinity of WT at intermediate acetylation ($\pAc = 0.5$), the BD1--Ac well depth in \mBD2 is increased by a factor of 1.131 relative to WT.
All non-chromatin interactions are otherwise unchanged.
A quantitative description of this matching procedure is provided in the caption of \figref{fig:4S3}.
Differences between WT and \mBD2 are therefore interpreted as arising primarily from the reduction in active binding valency, with the effective affinity matched at intermediate acetylation.

\subsection{Combinatorial binding model for multivalency}

To understand how multivalency sharpens the dependence of co-condensation on acetylation density, we construct a combinatorial model of BRD4 binding to a nucleosome with multiple acetylatable histone tails. 
We seek an expression for the configurational partition function $Z$ describing multivalent BRD4 binding to an acetylated chromatin segment.
The segment contains $\Nac$ nucleosomes, each of which has $\ns$ acetylatable histone tails.
Acetylation occurs independently at the level of individual tails with probability $\pAc$, so $\Nac$ sets the size of the acetylated region, while $\pAc$ controls the distribution of available binding sites on each nucleosome.
Each BRD4 molecule contains $\nd$ active Ac-binding bromodomains, which determines the maximum number of simultaneous BD--Ac binding interactions in which a BRD4 molecule can participate.
The full partition function $Z$ is constructed from contributions of individual nucleosomes, accounting for the combinatorial number of ways in which multivalent BRD4 molecules can occupy available acetylated sites.

\paragraph{Disorder-averaged configurational partition function.}
We first consider whether a single BRD4 molecule can bridge two nucleosomes.
The maximum separation at which such a configuration is geometrically possible is
\begin{equation}
\dHHinteract \approx \sgmHH + 2 D_{\text{g,H3}} + \frac{\Rgbrd}{\sqrt{2}},
\end{equation}
where $\sgmHH$ is the effective nucleosome size and $D_{\text{g,H3}}$ and $\Rgbrd$ are the gyration sizes of the H3 tail and BRD4, respectively.
Using $D_{\text{g,H3}} \approx \lAA \sqrt{37} \approx \SI{2.3}{nm}$ and $\Rgbrd \approx \SI{6}{nm}$ gives $\dHHinteract \approx \SI{19.4}{nm}$, comparable to the nucleosome spacing $\aHH \approx \SI{17}{nm}$.
However, forming two binding interactions at this distance requires highly specific geometries (alignment of two H3 tails and precise positioning of BRD4), leading to a substantial entropic penalty.
We therefore assume that the net free-energy cost of inter-nucleosome bridging is unfavorable and neglect these configurations in the analysis that follows in this section.
Under this approximation, nucleosomes can be treated as independent binding units. The total partition function then factorizes into a product of identical single-nucleosome contributions,
\begin{equation}
Z = \ZH^{\Nac},
\end{equation}
where $\ZH$ denotes the partition function for BRD4 binding to a single nucleosome.

We now consider a single nucleosome.
Each nucleosome has $\ns = 4$ acetylatable histone tails, each acetylated with probability $\pAc$.
For the analytic model, we approximate acetylation on each nucleosome as independent across tails with probability $\pAc$, which yields the binomial distribution
\begin{equation}
P_k = \binom{\ns}{k} \pAc^k (1-\pAc)^{\ns-k}.
\end{equation}
Because acetylation is quenched over a given chromatin realization, we use a disorder-averaged free energy
\begin{equation}
-\kT \ln \ZH = -\kT \sum_{k=0}^{\ns} P_k \ln z_k,
\end{equation}
where $z_k$ denotes the partition function of a nucleosome with $k$ available binding sites.
This corresponds to the disorder-averaged single-nucleosome partition function
\begin{equation}
\ZH = \prod_{k=0}^{\ns} z_k^{P_k},
\end{equation}
where $z_0 = 1$ by definition.

Given $k$ active sites, BRD4 molecules bind combinatorially.
Each molecule has $\nd$ active Ac-binding domains ($\nd=2$ for WT, $\nd=1$ for \mBD2).
We describe a class of binding configurations by a vector
\begin{equation}
\vec{f} = \{f_1, f_2, \dots, f_m\},
\end{equation}
where $m = |\vec{f}|$ is the number of bound molecules and $f_i$ is the number of domains used by molecule $i$.
The total number of occupied binding sites is
\begin{equation}
\sum_{i=1}^{m} f_i \le k.
\end{equation}
Binding via $n$ domains is characterized by an effective association volume $K_n$.
We approximate
\begin{equation}
K_n \approx K_1 \left(\frac{K_1}{v_0}\right)^{n-1},
\end{equation}
where $v_0$ is the effective interaction volume of a bound molecule.
This approximation neglects geometric differences between binding configurations.
It is justified by the fact that the size of BRD4 satisfies $\Rgbrd \gtrsim d_{\mathrm{site}}$, where $d_{\mathrm{site}}$ is the typical separation between acetylation sites on a nucleosome, so variations in stretching and entropic penalties between configurations are small.

We now enumerate configurations within a fixed filling vector $\vec{f}$.
A molecule binding through $f_i$ domains can choose which of its $\nd$ domains participate in $\binom{\nd}{f_i}$ ways.
Binding sites are assigned sequentially: when assigning molecule $i$, the previous molecules have already occupied $\sum_{q=1}^{i-1} f_q$ sites, so there are $k-\sum_{q=1}^{i-1} f_q$ available sites remaining, from which $f_i$ sites can be chosen.
This gives a factor $\binom{k-\sum_{q=1}^{i-1} f_q}{f_i}$.
For each molecule, the $f_i$ selected domains can be paired with the $f_i$ selected sites in $f_i!$ distinct ways.
If multiple molecules bind through the same number of domains, permuting those molecules does not generate a distinct filling.
We therefore divide by the multiplicity factor
\begin{equation}
R(\vec{f})! = \prod_a r_a!,
\end{equation}
where $r_a$ is the number of molecules in $\vec{f}$ with $f_i=a$.
We group fillings by the total number of occupied sites.
For a fixed total occupancy $j \le k$, we define $F_{\nd,j}$ as the set of distinct filling vectors $\vec{f}$ whose entries are positive integers no larger than $\nd$ and whose entries sum to $j$, with vectors related by permutation counted only once:
\begin{equation}
F_{\nd,j}=\left\{
\vec{f}=\{f_1,\dots,f_m\}:1 \le f_i \le \nd,\;\sum_{i=1}^{m} f_i = j
\right\}
/\,\text{permutations}.
\end{equation}

The disorder-averaged single-nucleosome partition function is then
\begin{align}
\label{eq:combinatorial_binding_model}
\ZH(c) &= \prod_{k=0}^{\ns} z_k(c)^{P_k}, \\
z_k(c) &= 1 + \sum_{j=1}^k \sum_{\vec{f} \in F_{\nd,j}}
\frac{1}{R(\vec{f})!}
\prod_{i=1}^{m}
(c K_{f_i}) f_i! \binom{\nd}{f_i}
\binom{k - \sum_{q=1}^{i-1} f_q}{f_i}.
\end{align}
Here, $z_k(c)$ sums over all distinct ways of occupying $k$ available binding sites with multivalent BRD4 molecules, including combinatorial choices of domains, binding sites, and their assignments, while $\ZH(c)$ averages this contribution over the distribution of acetylation states.

\paragraph{Adsorption isotherms.}
We now compute the expected occupancy of BRD4 on a nucleosome from the partition function. The mean number of bound molecules is obtained as
\begin{equation}
\nB(c, \dots) = c \frac{\partial \ln \ZH(c, \dots)}{\partial c}.
\end{equation}
For the single-valent case ($\nd=1$), this reduces to
\begin{equation}
\nB(c, \pAc, K_1) = 4 \pAc \frac{K_1 c}{1 + K_1 c},
\end{equation}
which has the form of a standard Langmuir adsorption isotherm.
Here the factor $4\pAc$ reflects the mean number of acetylated sites per nucleosome.

For the bivalent WT case ($\nd=2$), a closed-form expression can be obtained for the simplified case $\ns=4$ with identical effective single-domain association volume $K_1$ and a single effective two-domain association volume $K_2$, neglecting distinctions between H3/H4 tails and BD1/BD2 identities. Under these assumptions, the expression includes terms up to $(\pAc)^4$:
\begin{equation}
\begin{aligned}
    & \nB(c, \pAc, K_1, K_2) = 4\pAc \frac{2 K_1 c}{1 + 2 K_1 c} + \\
                        + & (4\pAc)^2 \frac{-3 (K_2 c) (2 K_1 c - 1)}{4 (1 + 2 K_1 c) ((1 + 2 K_1 c)^2 + 2 K_2 c)} + \\
                        + & (4\pAc)^3 \frac{3 (K_2 c)^2 (2 K_1 c - 1)}{2(1 + 2 K_1 c)((1 + 2 K_1 c)^4 + 8 K_2 c (1 + 2 K_1 c)^2 + 12 (K_2 c)^2)} + \\
                        + & (4\pAc)^4 \frac{-3 (K_2 c)^3 (2 K_1 c - 1)(5 + 20 K_1 c + 6 K_2 c + 20 (K_1 c)^2)}{8(1 + 2 K_1 c)((1 + 2 K_1 c)^8 + 20(K_2 c)(1 + 2 K_1 c)^6 + 120(K_2 c)^2(1 + 2 K_1 c)^4 + 240 (K_2 c)^3 (1 + 2 K_1 c)^2 + 144 (K_2 c)^4 )}.
\end{aligned}
\end{equation}

\paragraph{Consequences for co-condensation sensitivity to acetylation density.}
The effect of multivalency arises from the combinatorial growth in the number of binding configurations.
For $\nd=1$, the number of configurations scales linearly with the number of available sites $k$.
For $\nd=2$, configurations scale combinatorially; for example,
\begin{equation}
\binom{4}{2} = 6 > \binom{4}{1} = 4.
\end{equation}
As a result, highly acetylated nucleosomes contribute disproportionately to the partition function when multivalent binding is allowed.
The enhancement becomes significant at the smallest $k$ for which multivalent configurations outnumber single-site configurations,
\begin{equation}
k^*(\nd) = \min\left\{k : \binom{k}{\nd} > \binom{k}{1}\right\},
\end{equation}
which gives $k^* = 4$ for $\nd=2$. Since $\ns = 4$, this threshold coincides with the maximal acetylation state of a nucleosome.

This has direct consequences for co-condensation.
At low acetylation density $\pAc$, nucleosomes with $k \ge k^*$ are rare, and binding is dominated by effectively single-valent interactions.
As $\pAc$ increases, the probability of highly acetylated nucleosomes rises sharply, and these sites contribute more significantly to the binding free energy due to the combinatorial multiplicity of multivalent configurations.
This produces a nonlinear increase in the effective binding affinity with $\pAc$, which in turn sharpens the dependence of BRD4 recruitment and co-condensation on acetylation density for WT relative to \mBD2.
This mechanism is consistent with the observation of a threshold-like increase in BRD4 assembly and co-condensation with increasing acetylation density.

\subsection{Validation of combinatorial binding model using high-tension simulations}

\begin{figure*}
    \centering
    \includegraphics[width=0.8\textwidth]{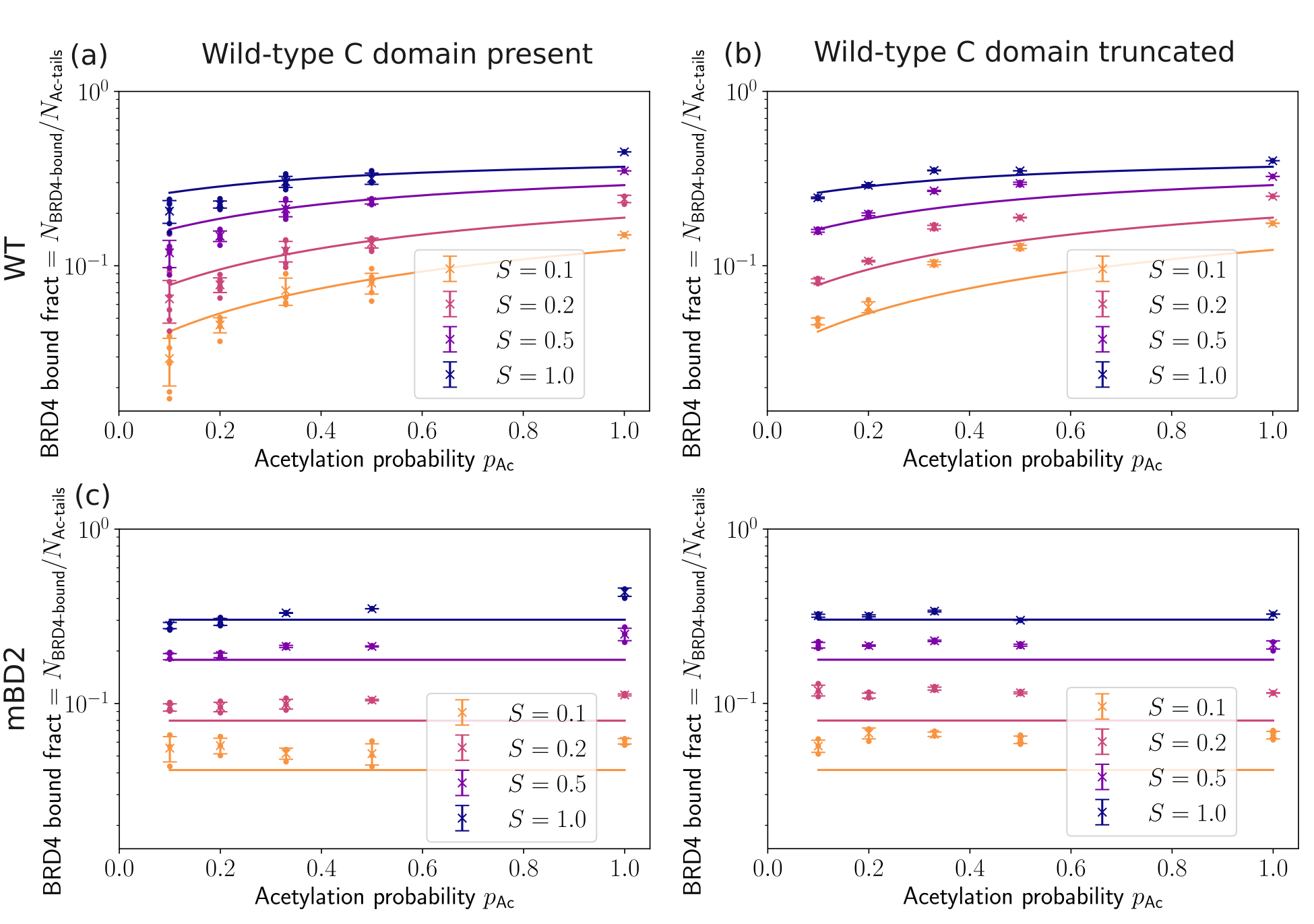}
    \caption{Validation of the combinatorial binding model in \eqref{eq:combinatorial_binding_model} by comparison to high-tension adsorption simulations. The plotted quantity is the number of bound BRD4 molecules normalized by the total number of available binding sites, $\nB(S,\pAc)/(4\pAc)$, as a function of acetylation density $\pAc$ and saturation $S$.
(a)~WT BRD4 with the C-terminal domain present. WT BRD4 shows a strong increase in normalized occupancy with $\pAc$, especially at low saturation, consistent with combinatorial multivalency.
(b)~WT BRD4 with the C-terminal domain truncated. Removing the C-terminal domain suppresses BRD4--BRD4 attraction and isolates the adsorption contribution.
(c)~Single-active-domain \mBD2 with the C-terminal domain present. The \mBD2 variant (see \figref{fig:4S3}) is constructed to match the effective histone tail--BRD4 binding affinity at intermediate acetylation ($\pAc \approx 0.5$), but shows little dependence on $\pAc$ at low saturation because it lacks multivalent chromatin binding. At higher saturation, \mBD2 shows an apparent increase with $\pAc$.
(d)~Single-active-domain \mBD2 with the C-terminal domain truncated. Comparison between (c) and (d) shows that the high-$S$ increase in \mBD2 occupancy arises from cooperative BRD4--BRD4 interactions mediated by the C-terminal domain rather than multivalent chromatin binding.}
    \label{fig:4S1}
\end{figure*}

To isolate the combinatorial binding contribution from co-condensation effects, we perform simulations at high chromatin tension, where polymer deformation and co-condensation are suppressed.
In this regime, BRD4 binding to chromatin is governed primarily by local binding interactions.
We therefore compare simulation results to the combinatorial binding model by measuring the equilibrium occupancy of BRD4 as a function of acetylation density $\pAc$ and saturation $S$ (\figref{fig:4S1}).
To remove trivial scaling with the number of available binding sites, we consider the normalized occupancy $\nB(S,\pAc)/(4\pAc)$.

At low saturation ($S \ll 1$), where BRD4--BRD4 interactions are negligible, the simulation results directly reflect the combinatorial binding mechanism.
In this regime, WT BRD4 exhibits a strong increase of normalized occupancy with $\pAc$, while \mBD2 shows little or no dependence on $\pAc$, consistent with the prediction that multivalency enhances the statistical weight of highly acetylated configurations.
At higher saturation ($S \sim 1$), both WT and \mBD2 exhibit an apparent increase in occupancy with $\pAc$.
However, this behavior is not captured by the combinatorial binding model alone and instead arises from cooperative BRD4--BRD4 interactions.
This is confirmed by simulations of the \mBD2-dC variant, which lacks the C-terminal domain responsible for these interactions and shows no such dependence on $\pAc$ at high $S$.

Together, these results demonstrate that the combinatorial binding model accurately describes BRD4--chromatin interactions in the regime where binding is dominated by local multivalency, while deviations at higher saturation arise from additional cooperative interactions not included in the combinatorial binding model.

\section{Kinetics and rate calculations}
\label{sec:kinetics}

\subsection{Diffusion-limited timescale normalization}

In the main text, rate densities are reported in diffusion-normalized units.
Here we define the saturation-dependent simulation diffusion time used for that normalization; the mapping of this timescale to physical units is described in \secref{sec:time_calibration}.
To compare assembly formation rates across different saturation conditions, we normalize time by a characteristic diffusion timescale for BRD4 in the dilute phase. We define
\begin{equation}
t_{\mathrm{Diff}}(S) = \frac{(2 R_{\rm g})^2}{6 D(S)},
\end{equation}
where $R_{\rm g}$ is the gyration radius of BRD4 measured in the dilute phase (and is approximately independent of $S$), and $D(S)$ is the center-of-mass diffusion coefficient of BRD4 at saturation $S$.
The length scale $2R_{\rm g}$ corresponds to diffusion over a molecular diameter.

\begin{figure*}
    \centering
    \includegraphics[width=1\textwidth]{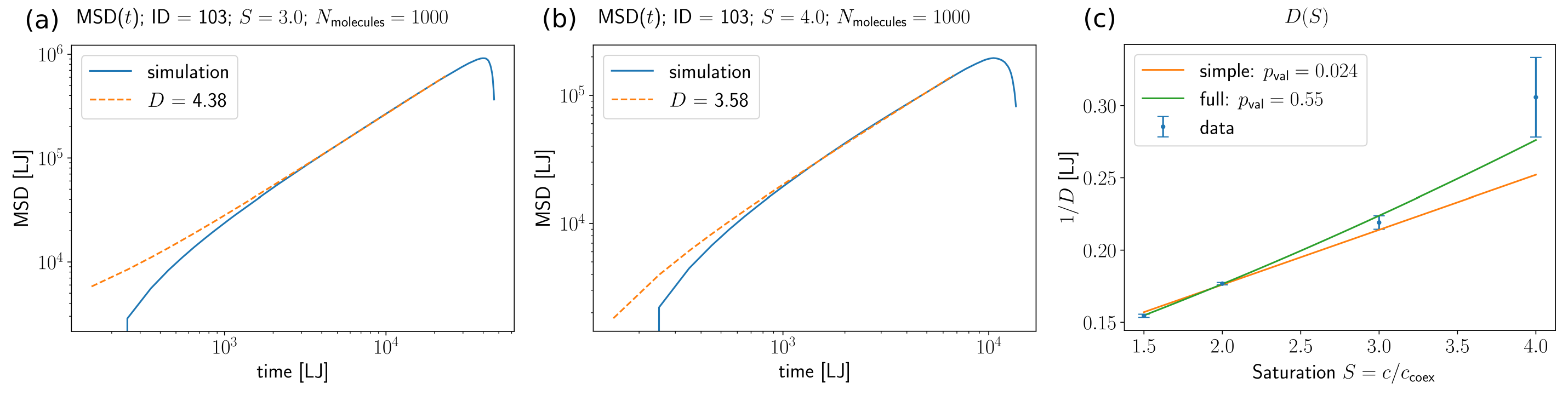}
    \caption{BRD4 self-diffusion and definition of the diffusion time scale.
(a)~Mean-squared displacement, $\mathrm{MSD}(t)$, of the BRD4 center of mass (CoM), averaged over 1000 BRD4 molecules at $S=3.0$. The downturn at long times reflects nucleation/assembly and is excluded from the diffusive fit.
(b)~Same as (a), but for $S=4.0$. Panels (a) and (b) show that the diffusive regime used to extract $D(S)$ is well resolved even when long-time nucleation affects the trajectory.
(c)~Diffusion coefficient $D(S)$ extracted from the linear diffusive regime of $\mathrm{MSD}(t)$ and fit to two models. The simple model assumes harmonic addition of Langevin and collision-limited contributions, $1/D_{\rm eff}(S)=1/\Dlangevin+1/D_{\rm kin}(S)=m/(\kT\tauLangevin)+S A_{\rm simple}/\sqrt{\kT}$, where increasing $S$ slows diffusion by increasing collision frequency. The full model additionally includes finite-size and slip-flow corrections, $D_{\rm eff}(c_{\rm far},N,L_z/L_{xy})=[m/(\kT\tauLangevin)+R_{\rm g}^{-1}\sqrt{m/\kT}(1+A_{\rm full}/(c_{\rm far}R_{\rm g}^3))^{-1}(1-B_{\rm full}(L_z/L_{xy})c_{\rm far}^{1/3}R_{\rm g}/N^{1/3})^{-1}]^{-1}$, where $c_{\rm far}$ is the far-field BRD4 concentration, $N$ is the number of BRD4 molecules, and $L_{xy}$ is the lateral box size. Here $A_{\rm simple}$, $A_{\rm full}$, and $B_{\rm full}$ are fit parameters. Using a goodness-of-fit threshold $\pval<0.05$ as the rejection criterion, the simple model is rejected whereas the full model is not. The full model is therefore used for estimating $D(S)$ across the complete range of simulated $S$ values.}
    \label{fig:5S1}
\end{figure*}

In our simulations, diffusion arises from two contributions: stochastic forcing from the Langevin thermostat and collisional interactions between particles.
Consistent with the model construction, these contributions are comparable in magnitude over the model parameter range of interest.
The effective diffusion coefficient $D(S)$ is obtained from the long-time slope of the mean-squared displacement of BRD4 molecules in the dilute phase, averaged over particles and trajectories (\figref{fig:5S1}).
We find that $D(S)$ decreases with increasing saturation due to the enhanced rate of collisional interactions.

Using the computed $D(S)$, we calculate $t_{\mathrm{Diff}}(S)$ and express all kinetic quantities in units of this timescale (\figref{fig:5S1}). Unless otherwise stated, $\tDiff$ denotes $t_{\mathrm{Diff}}(S)$ evaluated at the saturation of the corresponding simulation.
This normalization accounts for the saturation dependence of molecular diffusion and facilitates comparison between diffusion-limited and barrier-limited assembly regimes.

\subsection{Measurement of assembly formation rates}

\paragraph{Calculation of formation rate densities.}
Assembly formation rates are extracted from ensembles of independent simulations initialized without pre-formed assemblies.
For each trajectory, we define the first-passage time $\tFst$ as the earliest time at which the assembly size exceeds the threshold used to identify formation of a stable assembly.
The total formation rate for the simulated system is defined as
\begin{equation}
J = \frac{1}{\langle \tFst \rangle},
\end{equation}
where the average is taken over independent trajectories.
Throughout this section, $J$ denotes a total rate for the simulated system, whereas $j$ denotes a rate density obtained by normalizing $J$ by the appropriate nucleation volume.

To compare rates across different saturation conditions, we normalize by the characteristic interparticle spacing in the dilute phase.
Since $c_{\rm far} = S \ccoex$, the mean intermolecular distance scales as $c_{\rm far}^{-1/3} \sim S^{-1/3}$, and the characteristic diffusion time across this distance scales as $S^{-2/3}$.
We therefore define the dimensionless diffusion-normalized rate density
\begin{equation}
    \tilde{j}(S,\pAc)
    \equiv
    \frac{j(S,\pAc)\,t_{\rm Diff}(S)}
         {\ccoex S^{2/3}},
    \label{eq:jtilde}
\end{equation}
which factors out both the saturation-dependent molecular diffusion time and the expected concentration dependence associated with diffusion over the mean intermolecular spacing.

\paragraph{Calculation of off-chromatin rate densities.}
For off-chromatin nucleation, rates are measured in homogeneous simulations at $S>1$ and converted to volumetric rate densities,
\begin{equation}
j_{\mathrm{off}} = \frac{J_{\mathrm{off}}}{V}.
\end{equation}
These rates are fit using the classical nucleation theory (CNT) form
\begin{equation}
\ln \tilde{j}_{\mathrm{off}}(S)
=
A_{\mathrm{CNT}}
-
\frac{B_{\mathrm{CNT}}}{\ln^2 S},
\end{equation}
where $A_{\mathrm{CNT}}$ and $B_{\mathrm{CNT}}>0$ are fit parameters.
This expression reflects the expected CNT barrier scaling, $\Delta F^*/k_{\mathrm B}T \propto 1/\ln^2 S$, near coexistence.
This fit is used to interpolate, and where necessary cautiously extrapolate, $j_{\mathrm{off}}(S)$.

\paragraph{Calculation of on-chromatin rate densities.}
For on-chromatin assembly, the appropriate normalization volume is not simply the full simulation volume because binding sites are spatially localized on chromatin.
Approximating the relevant regions as cylindrical volumes around the chromatin fiber, the effective volume associated with the acetylated segment is
\begin{equation}
\label{eq:Vac}
\Vac = \Nac\, \aHH\, \pi \left(\frac{\DHist}{2} + \Rgbrdgas\right)^2,
\end{equation}
where $\aHH$ is the equilibrium nucleosome--nucleosome spacing, $\DHist$ is the effective nucleosome diameter, and $\Rgbrdgas$ is the dilute-phase BRD4 gyration radius.
The effective target volume $V_{\rm Ac}$ defined here is denoted by $v$ in the main text.
The corresponding volume excluded by non-acetylated chromatin is
\begin{equation}
\Vnonac = (\Nnucleosomes - \Nac)\, \aHH\, \pi \left(\frac{\DHist}{2} + \Rgtail\right)^2,
\end{equation}
where $\Rgtail$ is the characteristic tail size used to estimate the excluded region around non-acetylated chromatin.
Here $\Nnucleosomes$ is the total number of nucleosomes in the simulated chromatin fiber, and $V$ is the simulation volume.

The total measured rate in chromatin-containing simulations, $J$, includes both chromatin-enhanced nucleation near acetylated segments and background homogeneous nucleation in the remaining accessible volume.
We subtract the expected homogeneous contribution, $(V - \Vnonac)\joff$, and normalize by $\Vac$ to estimate the excess chromatin-associated rate density above the homogeneous background:
\begin{equation}
\jon =
\frac{J - (V - \Vnonac)\, \joff}{\Vac}.
\end{equation}
This construction provides an approximate chromatin-associated rate density while accounting for geometric exclusion and baseline homogeneous nucleation.

In regimes where assembly formation is not observed within the simulation time window, we report upper bounds on the total formation rate $J$ based on the total simulation time and number of independent trajectories, and convert these to the corresponding rate-density bounds using the same normalization conventions.

\subsection{Analysis of nucleation and assembly formation rates}

\paragraph{Analysis of nucleation rates above saturation ($S > 1$).}
In this regime, formation proceeds via barrier-limited nucleation of a critical cluster of size $N^*(S)$.
Consistent with CNT, the normalized rate depends strongly on supersaturation, with $\ln \tilde{j} = A - B/(\ln S)^2$ near coexistence \cite{porter2009phase}, as observed in both homogeneous simulations and in the presence of high-tension acetylated chromatin where co-condensation is suppressed.
Increasing the acetylation density of high-tension chromatin enhances local BRD4--chromatin interactions and lowers the effective nucleation barrier, leading to increased heterogeneous nucleation rates.
However, the dominant dependence on $S$ remains unchanged, indicating that the mechanism is still governed by barrier-limited nucleation.

\paragraph{Analysis of assembly formation rates below coexistence ($S < 1$).}
Below coexistence, classical nucleation theory predicts that nucleation of the bulk condensed phase is thermodynamically unfavorable.
However, in the presence of deformable chromatin, stable finite assemblies form via co-condensation.
Assembly formation rates are observed to vary weakly with saturation once acetylation exceeds the threshold required for stable co-condensation.
The diffusion-normalized rates $\tilde{j}$ collapse across saturation conditions, indicating that assembly kinetics are controlled primarily by diffusive transport rather than by a large nucleation barrier.

This behavior reflects a qualitatively different mechanism from nucleation-limited assembly.
In the co-condensation regime, the dominant rate-limiting step is the diffusion of BRD4 molecules from the dilute phase to the chromatin region.
Once bound, chromatin deformation and local stabilization proceed relatively rapidly, enabling growth without the need to overcome a large critical nucleus.
We note that this does not imply the complete absence of barriers.
At sufficiently low saturation, the system exhibits hysteresis, indicating the presence of a finite free-energy barrier.
Rather, there exists a broad intermediate regime in which this barrier is sufficiently small that assembly formation appears approximately diffusion-limited, leading to the observed plateau in normalized assembly formation rates.

\paragraph{Comparison between multivalent and single-active-domain models.}
The effect of multivalency on assembly kinetics is assessed by comparing WT BRD4 with the single-active-domain \mBD2 variant.
Above coexistence ($S>1$), both models exhibit nucleation-limited kinetics with similar dependence on saturation, indicating that the dominant mechanism is barrier-limited nucleation.
Differences between the models primarily reflect differences in effective binding strength.
Below coexistence ($S<1$), multivalency plays a qualitatively different role.
WT BRD4 exhibits rapid assembly across a broad range of subsaturated conditions once acetylation exceeds the threshold for co-condensation, whereas \mBD2 shows both reduced formation rates and a narrower range of conditions supporting assembly.
These kinetic differences mirror the equilibrium behavior: multivalency enhances the sensitivity to acetylation density and expands the regime in which co-condensation is stable, thereby increasing both the accessibility and robustness of assembly under subsaturated conditions.

\paragraph{Dependence of formation rates on acetylation density.}
Assembly formation rates as a function of acetylation density $\pAc$ further illustrate the link between kinetics and equilibrium stability.
At high chromatin tension, where assembly proceeds via nucleation, formation rates increase continuously with $\pAc$, consistent with a progressive reduction of the nucleation barrier.
At low chromatin tension, where co-condensation is permitted, formation rates exhibit a threshold-like increase at a critical $\pAc$ and then plateau over a broad range of subsaturation BRD4 concentrations.
This plateau indicates a crossover to approximately diffusion-limited assembly and reflects the onset of stable co-condensation.
For the single-active-domain \mBD2 model, this threshold is shifted and the plateau is less pronounced, with formation rates more strongly suppressed at low saturation.
This is consistent with the reduced ability of \mBD2 to stabilize co-condensed assemblies.
Overall, these results show that the dependence of assembly formation rates on $\pAc$ directly tracks the onset and strength of co-condensation, linking kinetic accessibility to the same multivalency-driven mechanisms that govern equilibrium stability.

\subsection{Absolute assembly timescales}
\label{sec:absolute_timescales}

\begin{figure*}
    \centering
    \includegraphics[width=1\textwidth]{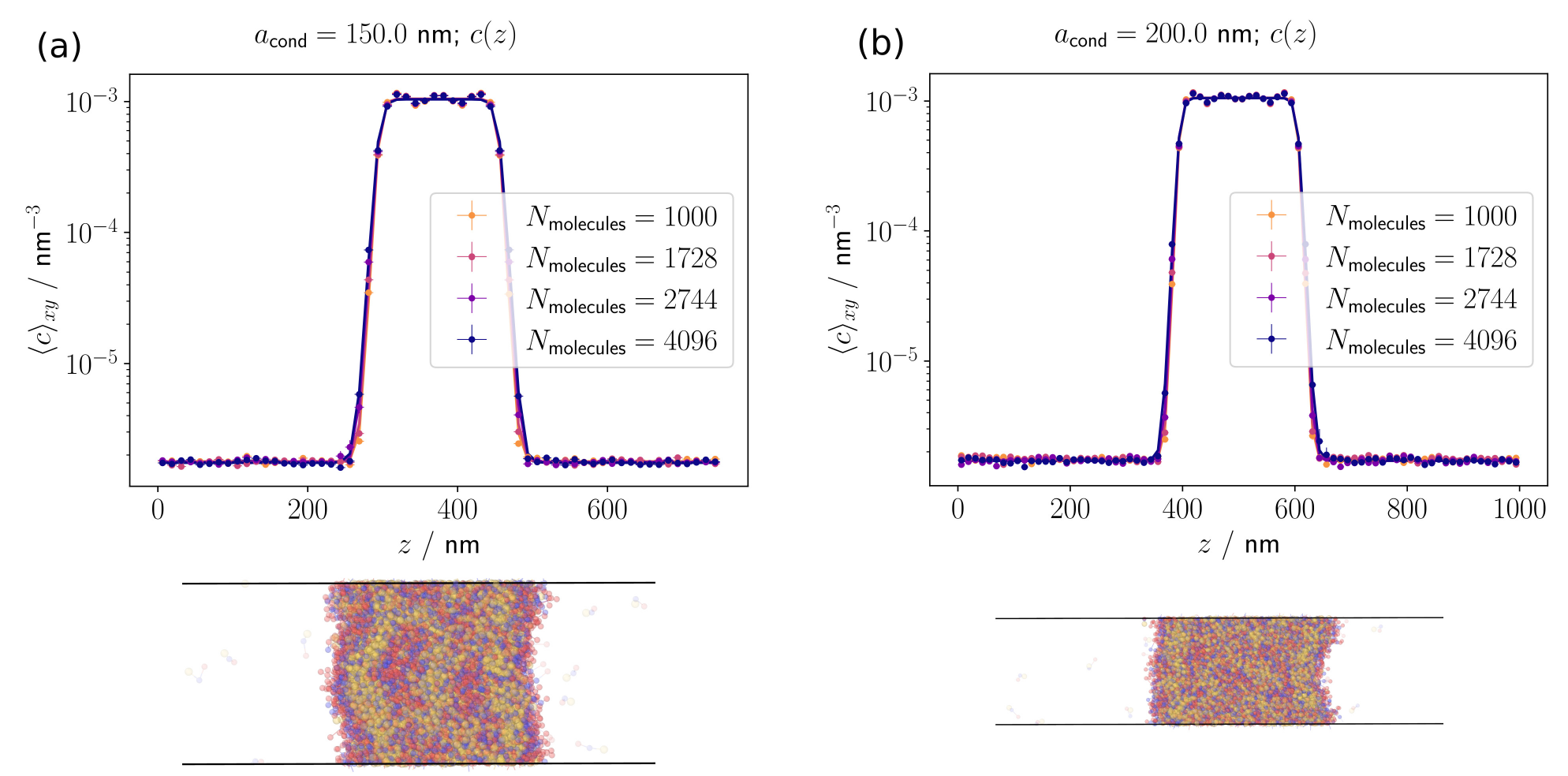}
    \caption{Systematic analysis of the influence of finite-size effects on simulated phase coexistence.
\textit{Top:} BRD4 concentration profiles in the $z$ direction, $c(z)$.
\textit{Bottom:} Snapshots of BRD4 coexistence simulations projected onto the $xz$ plane.
Left and right panels correspond to condensate widths of (a) $a_{\rm cond} = \SI{150}{nm}$ and (b) $a_{\rm cond} = \SI{200}{nm}$, respectively.
The overlap of density profiles for different numbers of BRD4 molecules (corresponding to different cross-sectional areas in the $xy$ plane) indicates the absence of significant finite-size effects.}
    \label{fig:1S3}
\end{figure*}

We finally use the diffusion calibration established in \secref{sec:time_calibration} to estimate physical assembly-formation times.
Using the dimensionless diffusion-normalized rate density defined in \eqref{eq:jtilde}, the corresponding physical rate density is
\begin{equation}
    j(S,\pAc)
    =
    \tilde{j}(S,\pAc)\,
    S^{2/3}
    \frac{\ccoex}{t_{\rm Diff}^{\rm exp}}.
    \label{eq:j_physical}
\end{equation}
From the bulk coexistence simulations, we obtain $\ccoex \approx \SI{1744\pm87}{\mu m^{-3}}$ (\figref{fig:1S3}), while \eqref{eq:tdiff_exp} gives $t_{\rm Diff}^{\rm exp}\approx\SI{4.4\pm0.3}{\mu s}$.
Thus,
\begin{equation}
    j
    \approx
    \SI{400\pm30}{\mu m^{-3} \mu s^{-1}} \!\times 
    \,\tilde{j}\,S^{2/3}.
\end{equation}

In the co-condensation regime, the normalized formation rates approach a plateau of approximately $\tilde{j}_{\rm plateau}\sim10^{-3}$, corresponding to
\begin{equation}
    j_{\rm plateau}
    \sim
    \SI{0.4}{\mu m^{-3} \mu s^{-1}} \!\times S^{2/3}.
\end{equation}
To convert this rate density into an assembly-formation time for a specific acetylated chromatin region, we use the same effective target volume $\Vac$ defined in \eqref{eq:Vac}.
For $\Nac=20$, with $\aHH=\SI{17\pm3}{nm}$, $\DHist\approx\SI{10.6}{nm}$, and $\Rgbrdgas\approx\SI{6}{nm}$, the effective radial extent is $\DHist/2+\Rgbrdgas\approx\SI{11}{nm}$.
The mean assembly-formation time is therefore
\begin{equation}
    \tau_{\rm assembly}
    =
    \frac{1}{V_{\rm Ac}j_{\rm plateau}}
    \sim
    \SI{20}{ms} \times S^{-2/3}.
    \label{eq:absolute_assembly_time}
\end{equation}
At $S=0.1$, near the regime that provides optimal targeting in the main text, this gives
\begin{equation}
  \tau_{\rm assembly}\sim\SI{90}{ms}.
\end{equation}

These estimates should be interpreted as predicted first-passage times for the formation of a stable chromatin-associated assembly.
A direct quantitative comparison with published live-cell measurements is not possible because existing experiments additionally depend on the number of available target sites and on detection thresholds.
Nevertheless, the predicted subsecond assembly-formation time is compatible with the rapid, seconds-scale formation of BRD4 condensates observed in U2OS cells \cite{Strom2024}.

\clearpage

\section{Supplementary Figures}

\begin{figure*}
    \centering
    \includegraphics[width=1\textwidth]{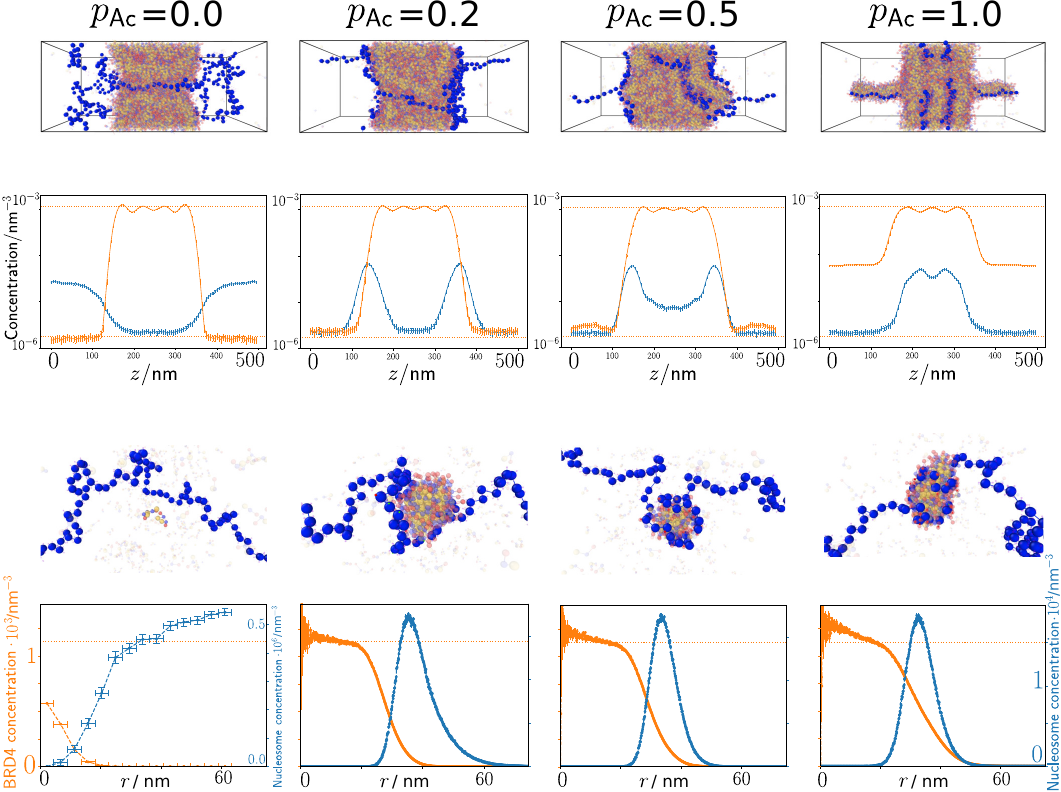}
    \caption{Flat BRD4 condensate interfaces partition chromatin outside, at the boundary, or within the bulk condensed phase depending on the acetylation density $\pAc$. Finite assemblies, by contrast, localize chromatin at their surface for a wide range of acetylation densities.
\textit{Left to right:} Increasing acetylation density $\pAc$ relative to the maximum possible number of acetylatable histone tails (i.e., all H3 and H4 histone tails, which constitute 50\% of all histone tails).
\textit{Top vs.\ bottom:} Approximately planar interfaces, corresponding to bulk condensates, versus finite assemblies. In simulations of bulk condensates, the entire chromatin fiber is acetylated with the same probability $\pAc$. In simulations of finite assemblies, a region of $\Nac = 20$ nucleosomes is acetylated with probability $\pAc$.
Top panels show concentration profiles in the $z$ direction of the corresponding systems, exploiting $\pm z$ symmetry.
Bottom panels show radially averaged concentration profiles.}
    \label{fig:2S0}
\end{figure*}

\begin{figure*}
    \centering
    \includegraphics[width=1\textwidth]{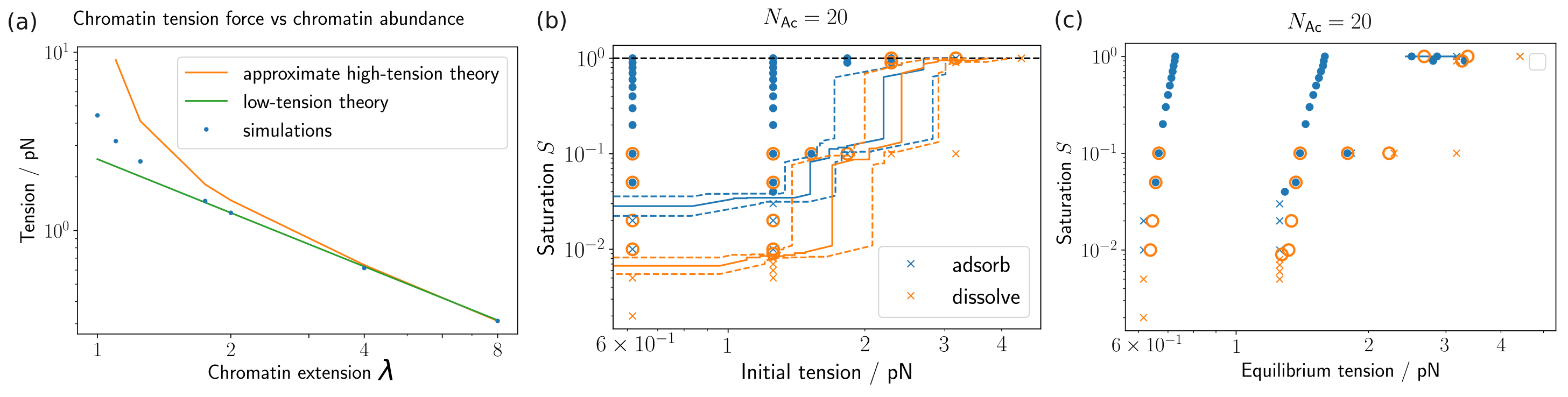}
    \caption{Chromatin tension calibration and finite-size tension shifts during assembly.
(a)~Relation between chromatin tension $\FT$ and chromatin extension $\lambda$. Blue points show chromatin-only simulations used to measure $\FT(\lambda)$; error bars are smaller than the symbol size. Reported tensions are obtained by linear interpolation of these data. Red and green curves show theoretical predictions: the red curve is the low-tension form $\FT/\kT=1/(\lambda a)$, with $a$ fitted to the three lowest-tension points, whereas the green curve is the nonlinear prediction for rigid nucleosome--nucleosome spacing, with the rigid step size fitted to the lowest-tension point.
(b)~Assembly phase diagram for $\Nac=20$, using the same assembly criterion as in the main text. The initial chromatin extension $\lambdainit$, prior to assembly, is used to determine the initial tension from (a). Solid and dashed lines show interpolated assembly boundaries and estimated uncertainty, respectively, for the indicated values of $\lambda$.
(c)~Same data as in (b), replotted using the post-assembly effective tension. The quantities $\lambdainit$ and $\lambdaeq$ denote the imposed initial extension and the effective extension after assembly formation, with corresponding tensions $\FTinit=\FT(\lambdainit)$ and $\FTeq=\FT(\lambdaeq)$. The difference between $\FTinit$ and $\FTeq$ is small at low tension but can become substantial near $\lambda\approx1$, where $\FT(\lambda)$ is highly nonlinear; this shift is a finite-size effect that depends on the accessible chromatin length.}
    \label{fig:3S1}
\end{figure*}

\begin{figure*}
    \centering
    \includegraphics[width=0.8\textwidth]{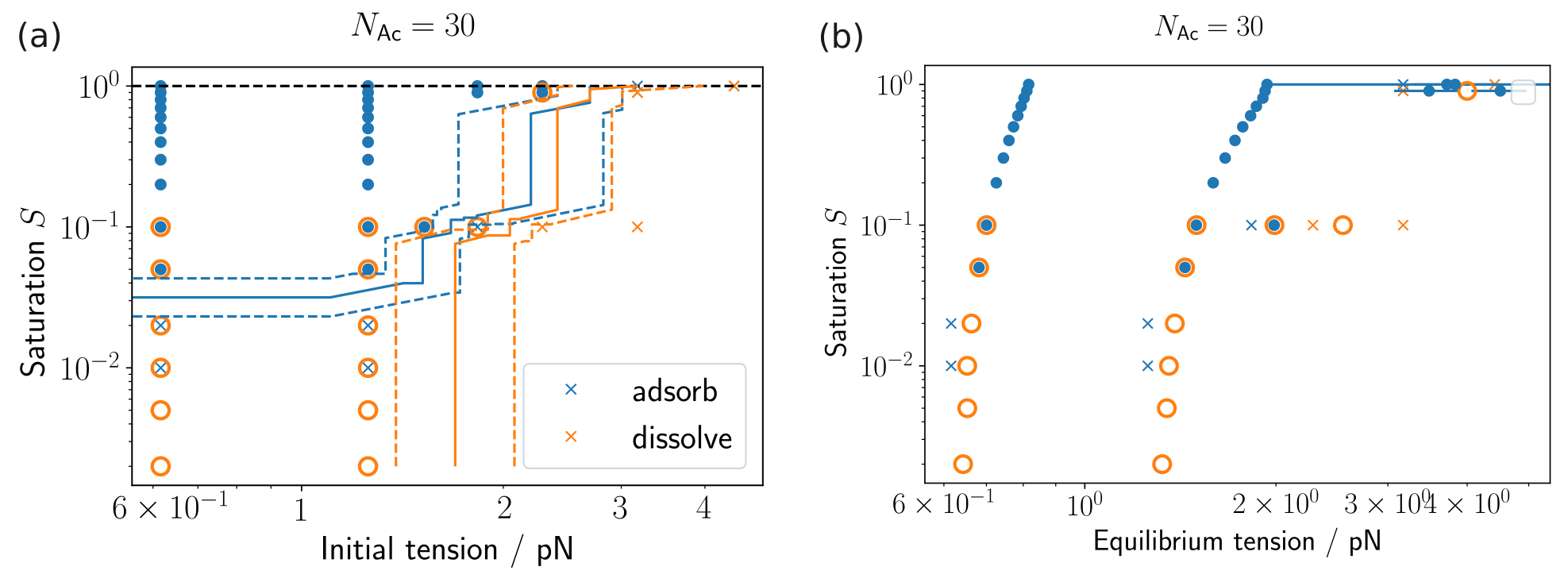}
    \caption{Data analogous to \figref{fig:3S1}b,c, but for $\Nac=30$ acetylated nucleosomes.
(a)~The assembly threshold is similar to the $\Nac=20$ case, indicating that the low-tension behavior is not strongly affected by equilibrium assembly size. The dissolution threshold shifts to lower $S$, consistent with increased metastable hysteresis for larger assemblies, but could not be determined directly due to slow simulation convergence at very low BRD4 concentration.
(b)~The shift from $\lambdainit$ to $\lambdaeq$ is larger than for $\Nac=20$, even at low $\FTinit$, because larger assemblies form at larger $\Nac$.}
    \label{fig:3S2}
\end{figure*}

\begin{figure*}
    \centering
    \includegraphics[width=0.75\textwidth]{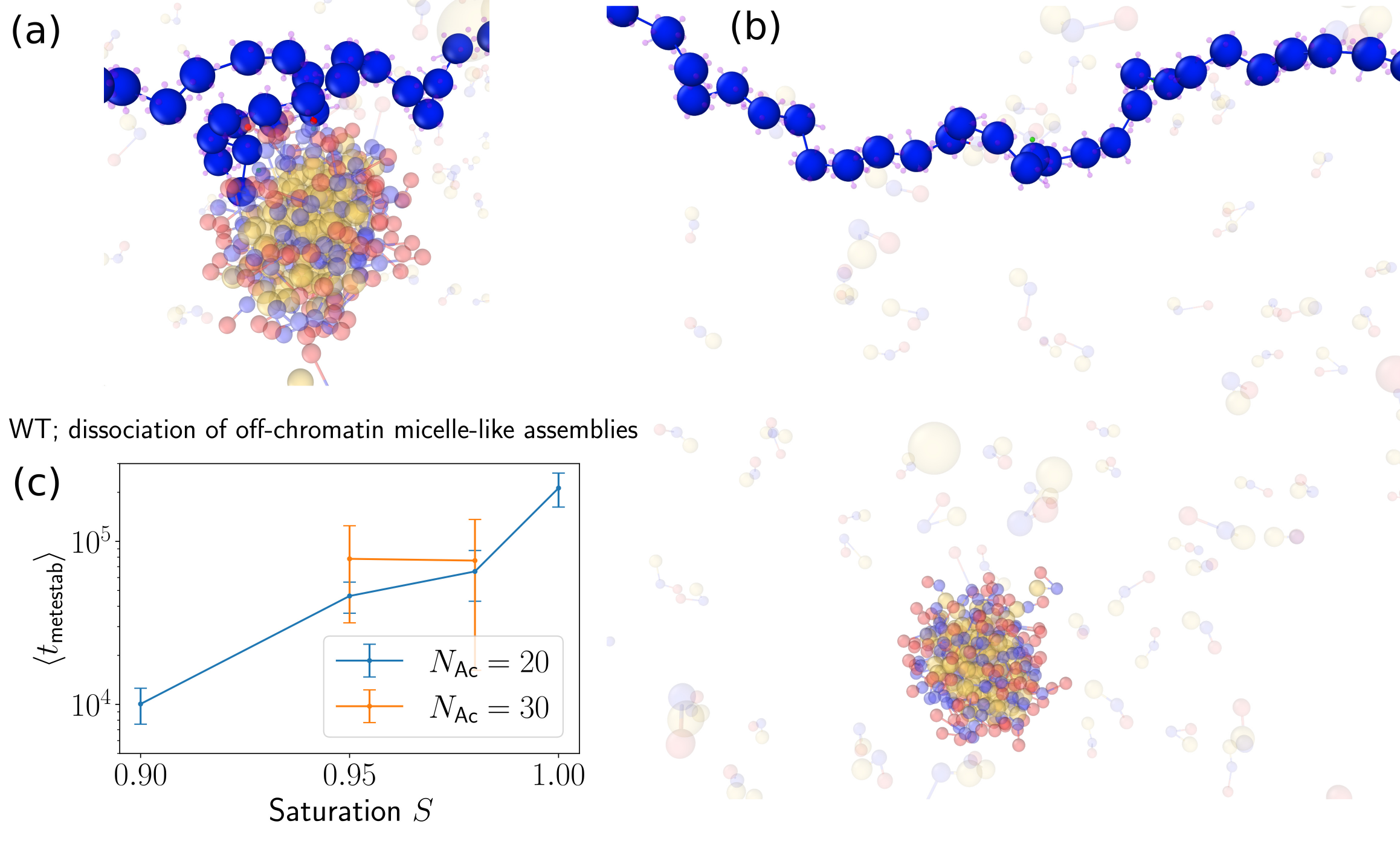}
    \caption{Micelle-like BRD4 assemblies near the saturation concentration.
(a)~Example configuration at $S=1.0$ after transiently increasing the range and strength of Ac-tail interactions to seed BRD4 assembly at multiple initially separated Ac sites, followed by restoration of the physical force field. The snapshot is taken after BRD4 molecules held only by artificial interactions have dissociated, but before substantial relaxation under the UCG force field.
(b)~The same system after many BRD4 self-diffusion times. Ac-mediated interactions are insufficient to maintain chromatin attachment, but the BRD4 assembly remains stable in solution, consistent with stabilization of micelle-like structures by the cone-like geometry of BRD4.
(c)~Mean metastable lifetime $\langle t_{\rm metastab}\rangle$ of micelle-like assemblies as a function of saturation $S$, where $t_{\rm metastab}$ is the time until dissolution of the initially formed assembly.
      }
    \label{fig:2S2}
\end{figure*}

\begin{figure*}
    \centering
    \includegraphics[width=0.8\textwidth]{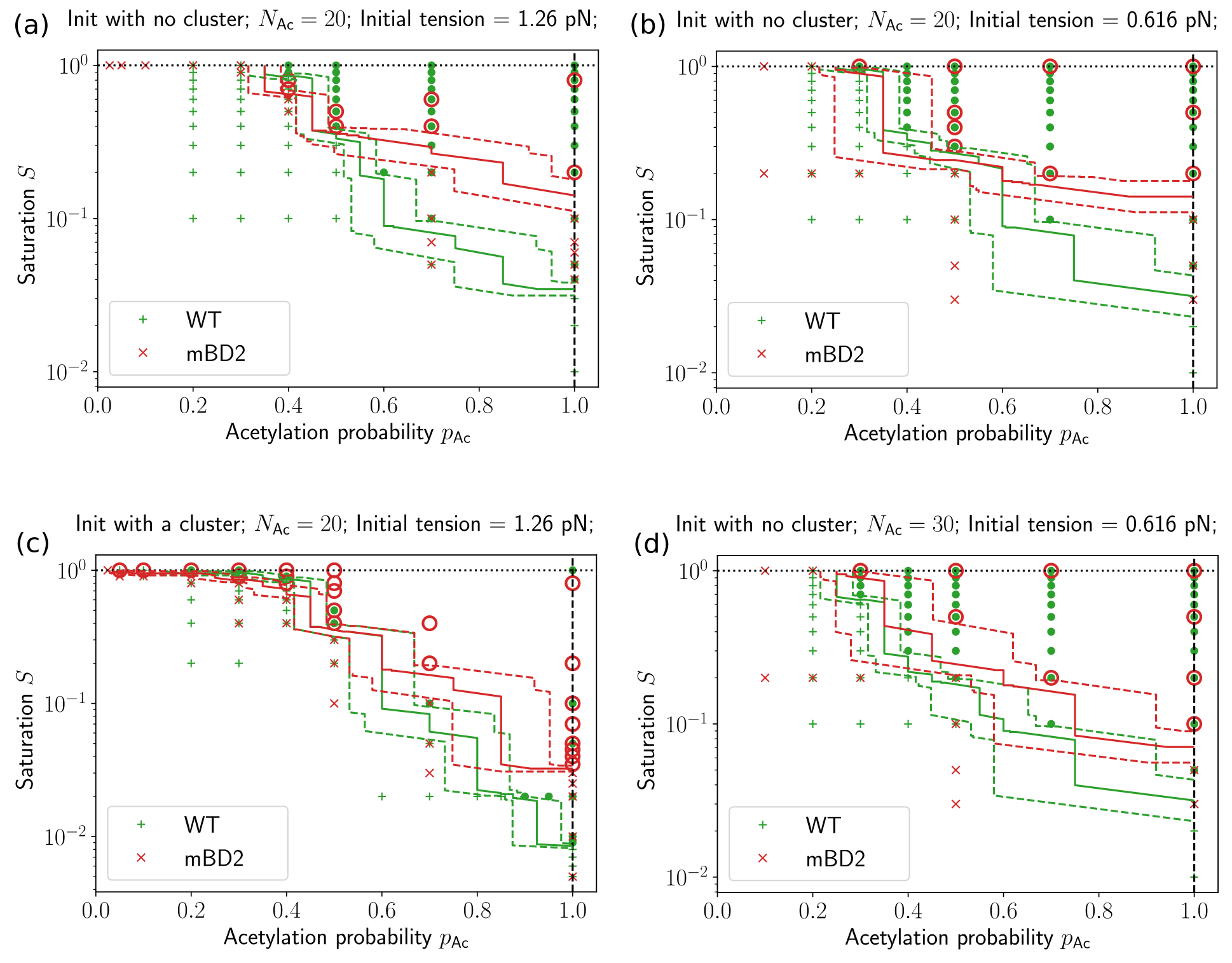}
    \caption{Robustness of acetylation sensitivity to tension, hysteresis, and acetylated-region size.
(a)~Assembly boundaries for the main-text conditions shown in \figref{fig:4}, obtained by linear interpolation.
(b)~Assembly boundaries at lower chromatin tension. The $\pAc$-sensitivity of \mBD2 is slightly reduced at high $\pAc$, but the effect is modest.
(c)~Dissolution boundaries for pre-existing assemblies under the same conditions as in (a). Hysteresis reduces the WT--\mBD2 gap at high $\pAc$, but WT remains more sensitive to acetylation. The metastable tail at low $\pAc$ and high $S$ arises from micelle-like assembly stabilization (see \figref{fig:2S2}).
(d)~Assembly boundaries at lower chromatin tension and larger acetylated-region size. Increasing $\Nac$ reduces the WT--\mBD2 difference, but WT remains more sensitive to acetylation.}
    \label{fig:4S2}
\end{figure*}

\begin{figure*}
    \centering
    \includegraphics[width=1\textwidth]{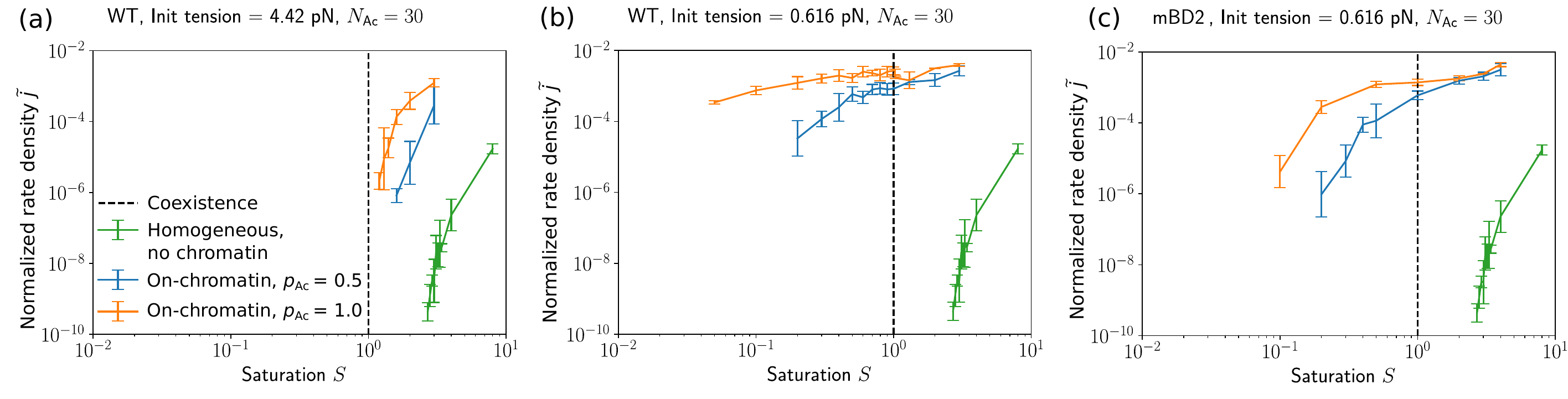}
    \caption{Data similar to those in \figref{fig:5}, but for $\Nac = 30$ acetylated nucleosomes. The qualitative conclusions are similar. The main difference is that the plateau effect is more pronounced here than at $\Nac = 20$.}
    \label{fig:5S2}
\end{figure*}

\begin{figure*}
    \centering
    \includegraphics[width=1\textwidth]{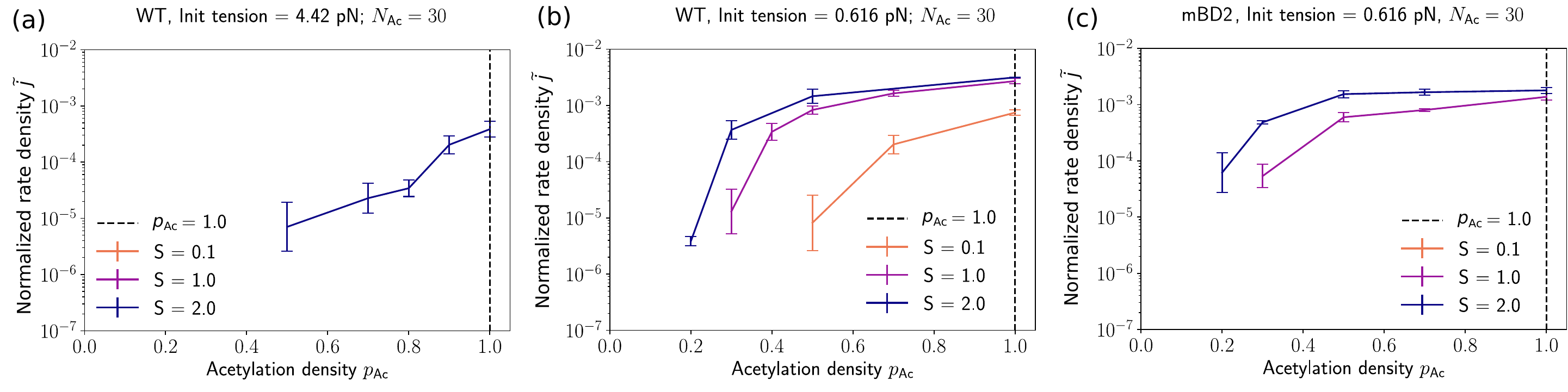}
    \caption{Data analogous to \figref{fig:6}, but for $\Nac = 30$. The conclusions are similar, and the effect of $\Nac$ is consistent with the comparison between \figref{fig:5} and \figref{fig:5S2}, whereby larger $\Nac$ leads to a broader rate plateau.}
    \label{fig:6S1}
\end{figure*}

\end{document}